\documentclass[fleqn,usenatbib]{mnras}
\usepackage{newtxtext,newtxmath}
\usepackage[T1]{fontenc}
\usepackage{ae,aecompl}
\usepackage{graphicx}	
\usepackage{amsmath}
\usepackage{times}
\usepackage{verbatim}
\usepackage{subfigure}
\usepackage{float}
\usepackage{color}

\usepackage{amssymb}

\title[Cold quenched galaxies in TNG]{Quenched, bulge-dominated, but dynamically cold galaxies in IllustrisTNG and their real-world counterparts}

\author[S. Lu et al.]
{Shengdong Lu$^{1}$\thanks{E-mail: \url{lushengdong@tsinghua.edu.cn}},
Dandan Xu$^{1}$\thanks{E-mail: \url{dandanxu@tsinghua.edu.cn}},
Sen Wang$^{1}$,
Yunchong Wang$^{2}$,
Shude Mao$^{1,3}$,
Xiaoyang Xia$^{4}$,
\and
Mark Vogelsberger$^{5}$,
Lars Hernquist$^{6}$
\\
\\
$^{1}$Department of Astronomy, Tsinghua University, Beijing 100084, China\\
$^{2}$Kavli Institute for Particle Astrophysics and Cosmology, Physics Department, Stanford University, Stanford, CA 94305, USA\\
$^{3}$National Astronomical Observatories, Chinese Academy of Sciences, 20A Datun Road, Chaoyang District, Beijing 100101, China\\
$^{4}$Tianjin Astrophysics Center, Tianjin Normal University, Tianjin 300387, China\\
$^{5}$Kavli Institute for Astrophysics and Space Research, Department of Physics, MIT, Cambridge, MA 02139, USA\\
$^{6}$Harvard-Smithsonian Center for Astrophysics, 60 Garden Street, Cambridge, MA 02138, USA
}
\date{Accepted ***. Received ***; in original form ***}
\pubyear{2021}

\begin{document}
\label{firstpage}
\pagerange{\pageref{firstpage}--\pageref{lastpage}}
\maketitle
\begin{abstract}
Galaxy morphologies, kinematics, and stellar populations are thought to be linked to each other. However, both simulations and observations have pointed out mismatches therein. In this work, we study the nature and origin of the present-day quenched, bulge-dominated, but dynamically cold galaxies within a stellar mass range of $10.3\,\leqslant\,\log\,M_{\ast}/\mathrm{M_{\odot}}\,\leqslant\,11.2$ in the IllustrisTNG-100 Simulation. We compare these galaxies with a population of normal star-forming dynamically cold disc galaxies and a population of normal quenched dynamically hot elliptical galaxies within the same mass range. The populations of the present-day quenched and bulge-dominated galaxies (both being dynamically cold and hot) used to have significantly higher star-formation rates and flatter morphologies at redshift of $z\sim 2$. They have experienced more frequent larger mass-ratio mergers below $z \sim 0.7$ in comparison to their star-forming disc counterparts, which is responsible for the formation of their bulge-dominated morphologies. The dynamically cold populations (both being star-forming and quenched) have experienced more frequent prograde and tangential mergers especially below $z \sim 1$, in contrast to the dynamically hot ellipticals, which have had more retrograde and radial mergers. Such different merging histories can well explain the differences on the cold and hot dynamical status among these galaxies. We point out that the real-world counterparts of these dynamically cold and hot bulge-dominated quenched populations are the fast- and slow-rotating early-type galaxies, respectively, as seen in observations and hence reveal the different evolution paths of these two distinct populations of early-type galaxies.
\end{abstract}


\section{Introduction}
\label{sec:introduction}

The \textbf{N}ormal star-forming \textbf{D}isc (NDs) and \textbf{N}ormal quenched \textbf{E}lliptical (NEs), in this context refer to two distinct types of galaxies above the Milky-Way stellar mass scale and below the galaxy group and cluster scales. The former are classically described as star-forming rotating discs, each word in turn, depicting the star-forming activity, the dynamical status and morphology. Correspondingly, the latter population can be described as quenched hot-orbit dominated ellipticals. The two types of galaxies are also conventionally referred to as late-type and early-type galaxies, respectively, albeit the latter have been shown to be in a later evolution stage as some end products of galaxy mergers.    

Recent studies from both observational and theoretical perspectives have realized some ``mismatches'' among galaxy morphology, colour, stellar population, star-forming activity, and dynamical status, which are traditionally thought to be closely linked to each other. For example, \citet{Lu_et_al.(2021a)} studied the formation of (intermediate-mass) star-forming but dynamically hot disc galaxies in the IllustrisTNG Simulation~\citep{Marinacci_et_al.(2018),Naiman_et_al.(2018),Nelson_et_al.(2018),Nelson_et_al.(2019b),Pillepich_et_al.(2018b),Pillepich_et_al.(2019),Springel_et_al.(2018),Vogelsberger_et_al.(2018),Vogelsberger_et_al.(2020b)}, which captured the nature of the observed gas-star misaligned discs and counter-rotating galaxies (e.g., Chen et al. in prep.). At the more massive end, the simulation also produces quenched and bulge-dominated galaxies, which however remain dynamically cold, exhibiting co-rotation in their stellar kinematics~\citep{Xu_et_al.(2019)}. We refer to this type of galaxies as "CQs", namely the dynamically \textbf{C}old, but \textbf{Q}uenched (and bulge-dominated) galaxies.

Regarding the formation of early-type galaxies, the existing consensus is that major and minor mergers play a key role in transforming late-type galaxies into early-type galaxies. During galaxy mergers, the cold gas reservoirs are largely consumed or removed by the triggered starbursts and AGN activities, building up stellar bulges at centres of galaxies (e.g., \citealt{Barnes_et_al.(1992),Barnes_et_al.(1991),Barnes_et_al.(1996),Mihos_et_al.(1996), Barnes_et_al.(2002),Naab_et_al.(2003),Bournaud_et_al.(2005),Bournaud_et_al.(2007),Johansson_et_al.(2009a),Tacchella_et_al.(2016),Rodriguez-Gomez_et_al.(2017)}). With time, many recent observations of early-type galaxies have further come to infer a dichotomy within the population, i.e. the so-called ``fast rotators'' and ``slow rotators'', in particular those from the SAURON project (e.g. \citealt{Cappellari_et_al.(2007),Emsellem_et_al.(2007)}) and the ATLAS$^{\rm 3D}$ survey (e.g. \citealt{Emsellem_et_al.(2011), Cappellari_et_al.(2013)}). This dichotomy is also seen in the early-type galaxies in the IllustrisTNG Simulation (e.g., \citealt{Pulsoni_et_al.(2020),Pulsoni_et_al.(2021)}). From both the observational and the simulation perspectives, marked kinematic misalignments between the gas and stellar components seen in these galaxies have strongly indicated an external origin to their gas reservoirs (e.g., see \citealt{Davis_et_al.(2011)} and \citealt{Krajnovic_et_al.(2011)} for the ATLAS$^{\rm 3D}$ galaxies). This naturally links their formation paths to their merging and accretion environments and histories. Indeed, previous simulations revealed that large mass-ratio mergers with retrograde and radial incoming orbits may strongly influence the formation of dynamically hot slow-rotating early-type galaxies (e.g., \citealt{Bois_et_al.(2011),Li_et_al.(2018a)}), while gas-rich major mergers tend to result in fast-rotating oblate-shaped early-type galaxies (e.g, \citealt{Hoffman_et_al.(2010)}). Recently, \citet{Zeng_et_al.(2021)} also found that major mergers on radial orbits are more likely to increase the kinematic bulge-to-total ratios in galaxies.

In this work, we point out that the fast- and slow-rotating early-type galaxies are the real-world counterparts of the CQ and the NE populations in this study, respectively. It is theoretically intriguing how the CQ galaxy population have quenched their star formation and built up bulge-dominated morphologies while maintaining significantly co-rotating (i.e. dynamically cold) kinematic discs. To answer this question, we investigate and compare the evolution histories of the three above-mentioned galaxy populations, i.e., the CQs, NDs and NEs, that have stellar masses within the same range of $10.3\,\leqslant\,\log\,M_{\ast}/\mathrm{M_{\odot}}\,\leqslant\,11.2$. 
We find that in comparison to NDs, both CQs and NEs have experienced more frequent larger mass-ratio mergers at lower redshifts of $z\lesssim 0.7$. The dynamically cold galaxies (i.e., NDs and CQs) have experienced a dominant fraction of prograde and tangential mergers below $z\lesssim 1$ with a variety of merger mass ratios. NEs, on the other hand, have experienced the most frequent retrograde and radial mergers throughout the redshifts (also with various merger mass ratios). Such merger statistics can well explain the cold dynamical statuses in CQs (as well as in NDs), in comparison to the hot dynamical status among the NE galaxy population. 

This paper is organized as follows. In Section~\ref{sec:data}, we briefly introduce the IllustrisTNG-100 Simulation (Section~\ref{sec:tng}) and clarify the sample selection criteria of the galaxies in this study (Section~\ref{sec:sample}). In Section~\ref{sec:look_like}, we present the general properties of these dynamically cold quenched elliptical galaxies in terms of their morphologies, kinematics, and stellar population properties, in comparison to the normal star-forming disc sample and the normal quenched elliptical sample. In Section~\ref{sec:how_form}, we present the redshift evolution of key properties of the investigated galaxies (Section~\ref{sec:evol_paras}) and a statistical analysis of mergers these galaxies have experienced (Section~\ref{sec:merger}). In Section~\ref{sec:observation}, we present several key observational properties and predictions of the real-world counterparts to the simulated galaxies. Finally, conclusions and discussion are given in Section~\ref{sec:conclusion_and_discussion}.

\section{Methodology}
\label{sec:data}
\subsection{The IllustrisTNG Simulation}
\label{sec:tng}
\textit{The Next Generation Illustris Simulations} (IllustrisTNG, TNG
hereafter; \citealt{Marinacci_et_al.(2018),Naiman_et_al.(2018),Nelson_et_al.(2018),Nelson_et_al.(2019a),Pillepich_et_al.(2018b),Springel_et_al.(2018)}) are a suite of state-of-the-art magneto-hydrodynamic cosmological galaxy formation simulations carried out in large cosmological volumes with the moving-mesh code \textsc{arepo} \citep{Springel(2010)}. The IllustrisTNG Simulations have the same initial conditions as the original Illustris Simulations \citep{Genel_et_al.(2014),Vogelsberger_et_al.(2013),Vogelsberger_et_al.(2014b),Vogelsberger_et_al.(2014a),Nelson_et_al.(2015)}, but differ in the updated version of the galaxy formation model, such as the addition of ideal magneto-hydrodynamics, a new active galactic nucleus (AGN) feedback model \citep{Weinberger_et_al.(2017),Weinberger_et_al.(2018)}, and various modifications to the galactic winds, stellar evolution, and chemical enrichment schemes \citep{Pillepich_et_al.(2018a)}. Our readers are referred to \citet{Vogelsberger_et_al.(2020a),Vogelsberger_et_al.(2020b)} for reviews of the cosmological simulations. In this study, we use the full-physics version with a cubic box of $110.7\,\mathrm{Mpc}$ side length (TNG100), which has a mass resolution for baryonic and dark matter of $m_{\rm baryon}=1.4\times10^6\,{\rm M_{\odot}}$ and $m_{\rm DM}=7.5\times10^6\,{\rm M_{\odot}}$, respectively. The gravitational softening length for dark matter and stellar particles is $\epsilon_{\rm softening} = 0.5\,h\mathrm{^{-1}kpc}$. Galaxies in their host dark matter halos are identified using the {\sc subfind} algorithm \citep{Springel_et_al.(2001),Dolag_et_al.(2009)}. General galaxy properties have been calculated and publicly released by the TNG collaboration\footnote{\url{http://www.tng-project.org/data/}} \citep{Nelson_et_al.(2019a)}.

\subsection{Sample Selection}
\label{sec:sample}

\begin{figure*}
\includegraphics[width=1.5\columnwidth]{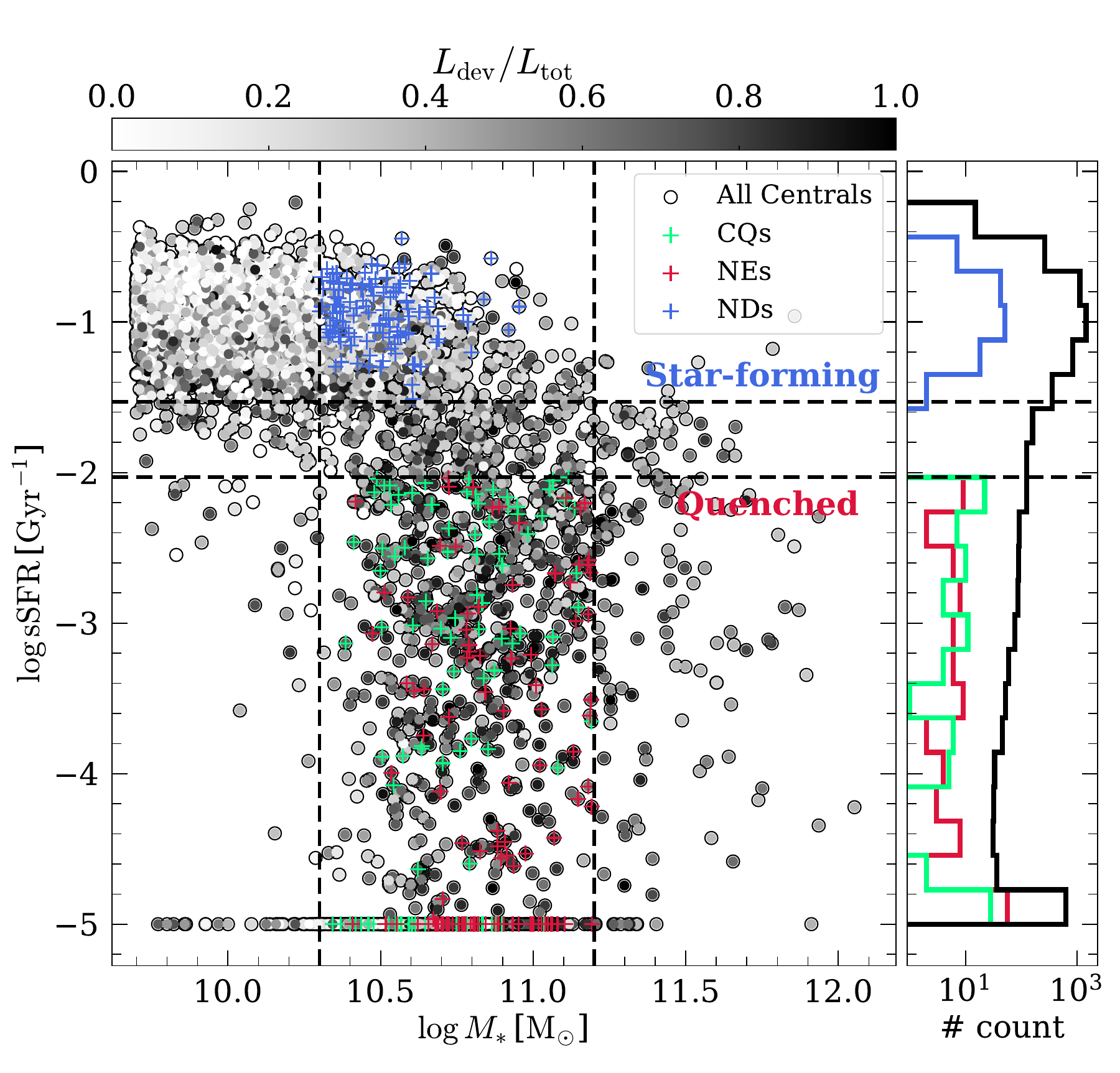}
\caption{Distributions of all the central galaxies on the $\log\,\mathrm{sSFR}-\log\,M_{\ast}$ plane, colour-coded by the de Vaucouleurs to total luminosity ratio, $L_{\rm dev}/L_{\rm tot}$. The two black dashed vertical lines indicate $\log\,M_{\ast}/\mathrm{M_{\odot}}=10.3$ and $\log\,M_{\ast}/\mathrm{M_{\odot}}=11.2$, between which our samples are selected. The two black dashed horizontal lines indicate $\log\,\mathrm{sSFR}/\mathrm{Gyr^{-1}}\approx -2$ (below which galaxies are defined to be quenched) and $\log\,\mathrm{sSFR}/\mathrm{Gyr^{-1}}\approx -1.5$ (above which galaxies are defined to be star-forming). The selected CQs, NEs, and NDs are indicated by green, red, and blue plus symbols, respectively. The histograms on the right show the distributions of $\log\,\mathrm{sSFR}$ of the total central sample (black), the selected CQs (green), NDs (blue), and NEs (red).}
\label{fig:all_galaxies}
\end{figure*}

\begin{figure*}
\includegraphics[width=1.8\columnwidth]{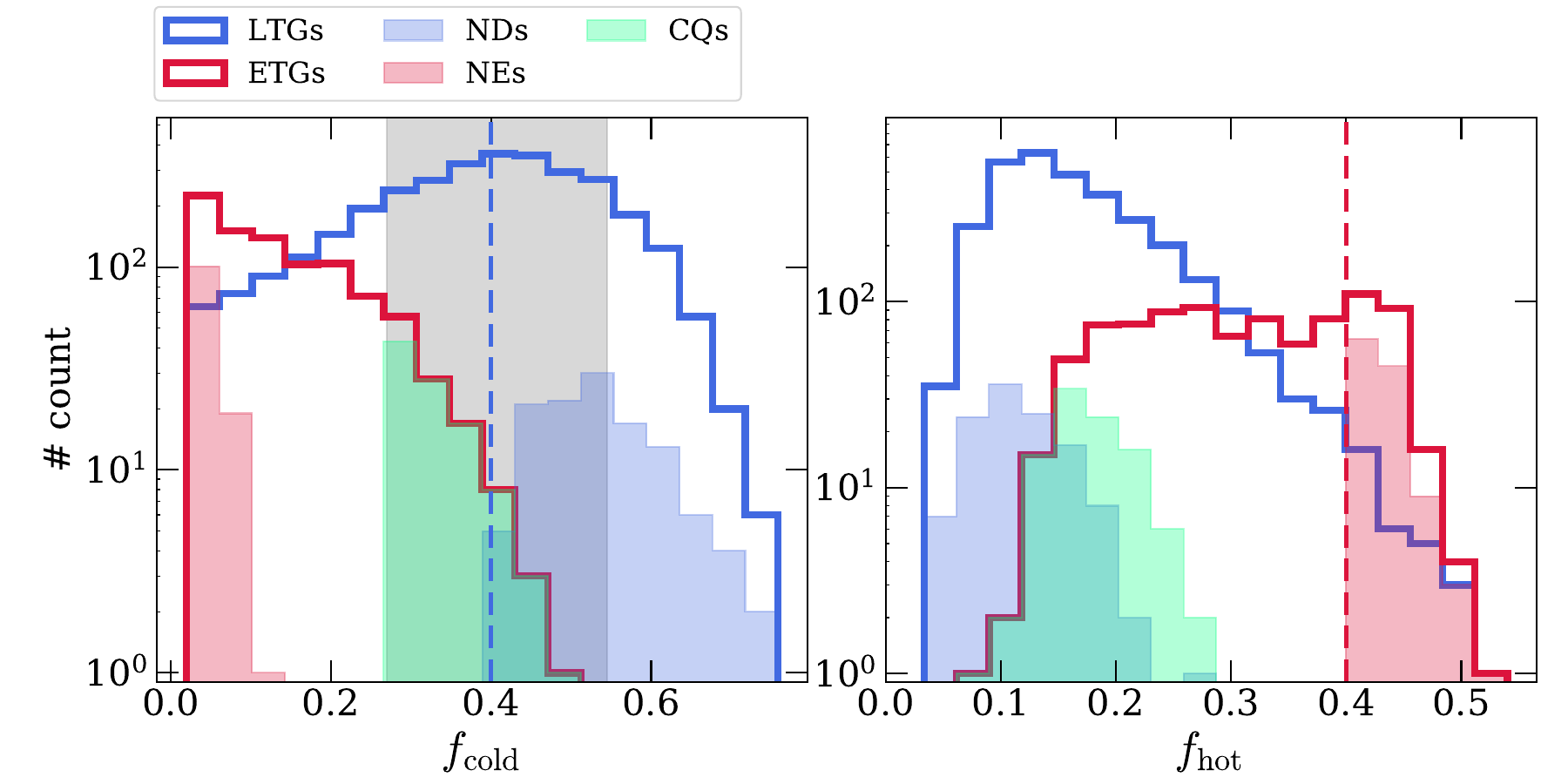}
\caption{Distributions of cold (left) and hot (right) orbital fractions. The distributions of early-type galaxies and late-type galaxies are shown by the red and blue hollow histograms, respectively. The distributions of CQs, NEs, and NDs are indicated by green, red, and blue shaded histograms, respectively. In the left panel, the grey shaded region is the 1$\sigma$ region of $f_{\rm cold}$ of late-type galaxies and the blue dashed line indicates $f_{\rm cold}=0.4$, above which NDs are randomly selected. In the right panel, the red dashed line indicates $f_{\rm hot}=0.4$, above which NEs are randomly selected.}
\label{fig:sample_forbit}
\end{figure*}

In this work, we first formulate an initial sample by only selecting central galaxies from $z=0$ that have stellar masses $M_{\ast} \geqslant\, 5\times 10^9 \mathrm{M_{\odot}}$ (where $M_{\ast}$ is the stellar mass of the galaxies within $\rm 30\,kpc$) in order to guarantee that the selected simulated galaxies are all well resolved. We then follow the practice of \citet{Lu_et_al.(2021a)} to select typical early- and late-type galaxies according to: (i) the specific star-formation rate (sSFR) and (ii) the de Vaucouleurs~\citep{de_Vaucouleurs(1948)} to total luminosity ratio, $L_{\rm dev}/L_{\rm tot}$\footnote{$L_{\rm dev}/L_{\rm tot}$ is obtained from fitting a two component model composed of a de Vaucouleurs and an exponential profile to the radial surface brightness distribution of the elliptical isophotes (see~\citealt{Xu_et_al.(2017)} for details).}. Specifically, the early-type galaxy sample shall have $\log\,\mathrm{sSFR}/\mathrm{Gyr^{-1}} \leqslant -2$ (i.e. quenched)\footnote{The numerical value of $-2$ is roughly 1 dex below the mean value of $\log\,\mathrm{sSFR}/\mathrm{Gyr^{-1}}$ for central galaxies with $\log\,M_{\ast}/\mathrm{M_{\odot}}<10.5$. See \citet{Lu_et_al.(2021a)} for more details.} and $L_{\rm dev}/L_{\rm tot}>0.5$ (i.e. bulge-dominated), while the late-type galaxy sample is required to satisfy $\log\,\mathrm{sSFR}/\mathrm{Gyr^{-1}}\geqslant -1.5$ (i.e. star-forming) and $L_{\rm dev}/L_{\rm tot}<0.5$ (i.e. disc-dominated).

Among the selected early- and late-type galaxies, we further classify and form three specific galaxy samples for this study, namely, normal star-forming disc galaxies, bulge-dominated quenched but dynamically cold galaxies, and normal (quenched and dynamically hot) elliptical galaxies, referred to as NDs, CQs and NEs, respectively. For all the three galaxy samples, we adopt a stellar mass range of $10.3\,\leqslant\,\log\,M_{\ast}/\mathrm{M_{\odot}}\,\leqslant\,11.2$ for the reason that both star-forming disc galaxies and quenched bulge-dominated galaxies can be found in this mass range; galaxies more massive than $10^{11.2}\,\mathrm{M_{\odot}}$ are no longer actively forming stars and are shown to have built up a substantial fraction of (ex-situ) stellar masses through major mergers (e.g., \citealt{Rodriguez-Gomez_et_al.(2017)}).

In order to quantify the dynamical status of galaxies, we use the instantaneous circularities $\lambda_z$ of stellar particles and define cold- and hot-orbit stars to have $\lambda_z>0.8$ and $|\lambda_z|<0.25$, respectively. The luminosity fractions $f_{\rm cold}$ and $f_{\rm hot}$ of stars in cold and hot orbits, respectively, are calculated within twice the stellar half mass radii, $2R_{\rm hsm}$ (see \citealt{Xu_et_al.(2019)} for the detailed definitions and calculations).

The CQ galaxy sample is composed of early-type (i.e. quenched and bulge-dominated) galaxies that have cold-orbit (i.e. $\lambda_z>0.8$) fractions $f_{\rm cold}>0.27$ such that the CQ galaxy sample shall have $f_{\rm cold}$ larger than the $1\sigma$ lower bound of that of the selected  {\it late-type} galaxies. This results in 100 CQs at $z=0$ from the simulation, making up $\sim 11\%$ of the total central early-type sample at $z=0$. To make a clear comparison for this study, we form the sample of NEs by randomly selecting extremely dynamically hot early-type galaxies (within the given stellar mass range and of roughly equal number) that have hot-orbit (i.e. $|\lambda_z|<0.25$) fractions $f_{\rm hot}>0.4$ and the sample of NDs with extremely dynamically cold late-type galaxies that have $f_{\rm cold}>0.4$.  

Table~\ref{table:table1} shows the basic sample statistics at $z=0$. The $\log\,\mathrm{sSFR}-\log\,M_{\ast}$ distributions of the three specific galaxy samples with respect to the overall central galaxy sample are shown in Fig.~\ref{fig:all_galaxies}. Histograms of the cold- and hot-orbit fractions $f_{\rm cold}$ and $f_{\rm hot}$ are given in Fig.~\ref{fig:sample_forbit}. As can be seen, early-type galaxies on average have lower $f_{\rm cold}$ and higher $f_{\rm hot}$ compared to late-type galaxies. The hot-orbit fractions in CQs are slightly higher than that of NDs and significantly lower than that of NEs; while our NEs typically have $f_{\rm cold}<0.1$, and NDs with $f_{\rm hot}<0.2$.

\begin{table}
\caption{Number of galaxies in different samples. Col.(1): all $z=0$ central galaxies with $M_{\ast} \geqslant\, 5\times 10^9\,\mathrm{M_{\odot}}$ from the IllustrisTNG Simulation; also the parent sample of Col.(2)$-$(6). Col.(2): early-type galaxies; Col.(3): late-type galaxies; Col.(4): selected cold quenched bulge-dominated galaxies (CQs); Col.(5): selected normal (hot) elliptical galaxies (NEs); and Col.(6): selected normal star-forming disc galaxies (NDs). See Section~\ref{sec:sample} for definition and selection criteria.} 
\setlength{\tabcolsep}{3mm}
\begin{tabular}{cccccc}
\hline
\hline
$N_{\rm central}$ & $N_{\rm early-type}$ & $N_{\rm late-type}$ & $N^{\rm sel}_{\rm CQ}$ & $N^{\rm sel}_{\rm NE}$ & $N^{\rm sel}_{\rm ND}$\\
\hline
5681 & 919 & 3177 & 100 & 121 & 120\\
\hline
\end{tabular}
\vspace{2mm}
\label{table:table1}
\end{table}

\section{What do they look like?}
\label{sec:look_like}
In this section, we present the general properties of the selected cold quenched (bulge-dominated) galaxies, normal (quenched and dynamically hot) ellipticals, and normal (star-forming) discs in terms of their morphologies (Section~\ref{sec:prop_morph}), stellar dynamics (Section~\ref{sec:prop_dynamics}), and stellar population properties (Section~\ref{sec:prop_sp}).

\subsection{Galaxy morphologies}
\label{sec:prop_morph}
\begin{figure*}
\includegraphics[width=2\columnwidth]{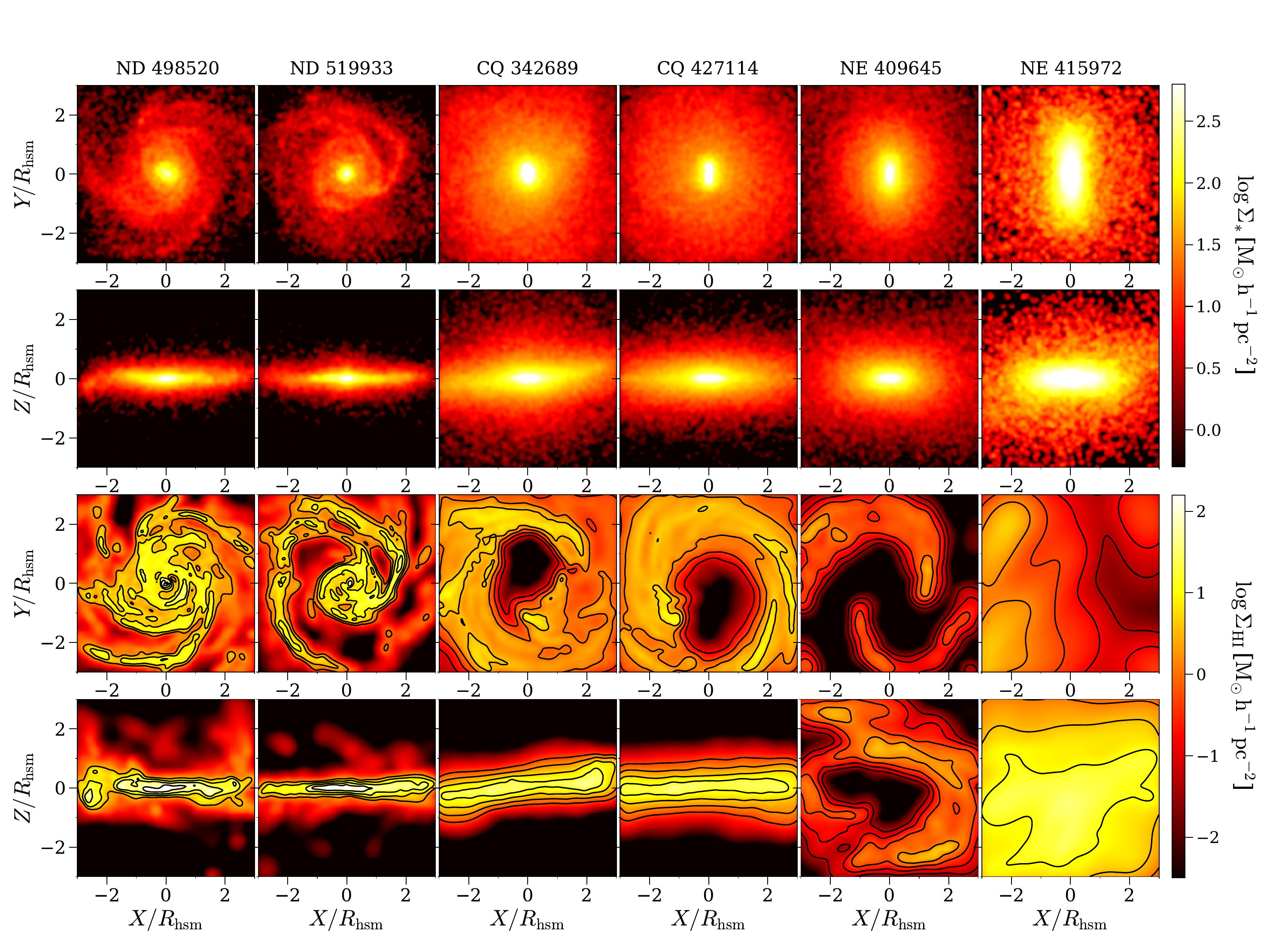}
\caption{Examples of two NDs (IDs: 498520, 519933), two CQs (IDs: 342689, 427114), and two NEs (IDs: 409645, 415972) from left to right. For each example, we show from the top to the bottom rows: (1) the face-on view of the stellar mass density map, (2) the edge-on view of the stellar mass density map, (3) the face-on view of the HI gas surface mass density map, and (4) the edge-on view of the HI gas surface mass density map. Contours of the HI surface mass densities are shown every 0.5 dex with a set of black curves.}
\label{fig:example_morphology}
\end{figure*}

\begin{figure*}
\includegraphics[width=1.8\columnwidth]{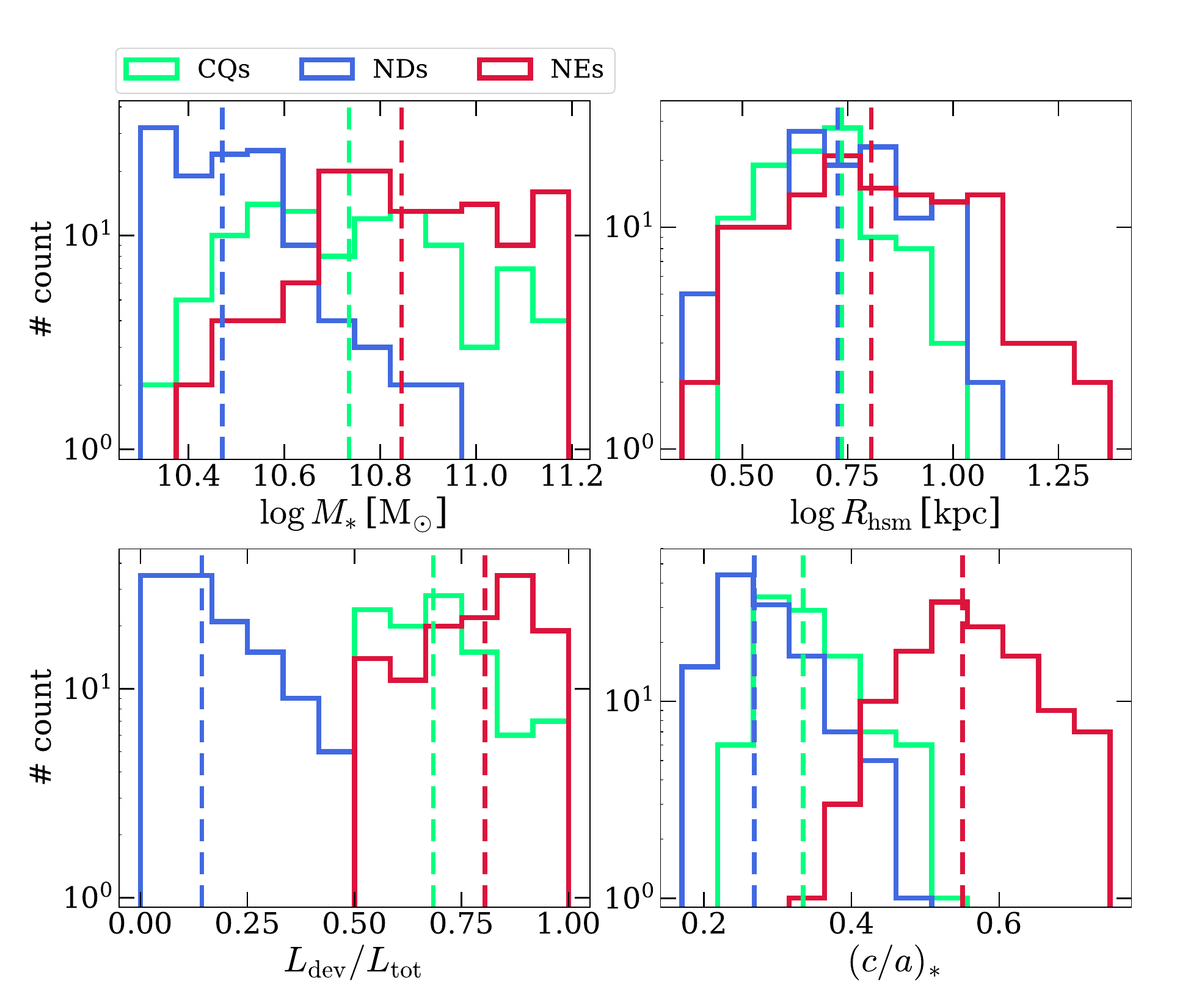}
\caption{Distributions of the stellar mass ($\log\,M_{\ast}$, top left), the half stellar mass radius ($\log\,R_{\rm hsm}$, top right), the de Vaucouleurs to total luminosity ratio ($L_{\rm dev}/L_{\rm tot}$, bottom left), and the shortest-to-longest principal axis ratio of stellar component ($(c/a)_{\ast}$, bottom right) for CQs (green), NEs (red), and NDs (blue). In each panel, the green, red, and blue dashed lines indicate the median value of the investigated property of CQs, NEs, and NDs, respectively.}
\label{fig:morphology}
\end{figure*}

In Fig.~\ref{fig:example_morphology}, we present the morphologies of both stellar and HI gas components for 6 examples (2 NDs, 2 CQs, and 2 NEs) from both the face-on and edge-on views. As expected, NDs (the first two columns) show clear spiral features from the face-on view in both stellar and gaseous components with thin stellar and gas discs from the edge-on view. NEs (the last two columns) show significantly thicker morphologies with irregular gas distributions. CQs, however, although being bulge-dominated, are seen to maintain relatively thinner stellar discs, compared to NEs. Besides, we find that CQs still host thin HI discs, which are in ring-like structures around the galaxy center. To better demonstrate the discrepancies in morphology between the three galaxy samples, we present, in Fig.~\ref{fig:morphology}, the distributions of several key properties which describe the morphologies of the stellar components of galaxies, namely: (1) the stellar mass $M_{\ast}$, (2) the stellar half mass radius $R_{\rm hsm}$, (3) the de Vaucouleurs to total luminosity ratio $L_{\rm dev}/L_{\rm tot}$, and (4) the shortest-to-longest axis ratio $(c/a)_{\ast}$. As can be seen, the three types of galaxies span slightly different stellar mass ranges: the $M_{\ast}$ distribution of NDs peaks at a slightly lower mass while that of NEs peaks at a higher mass. In comparison, CQs have a somewhat flatter distribution. Their stellar sizes $\log\,R_{\rm hsm}$ do not show obvious differences except that NEs have a tail to larger sizes. By sample construction, the distribution of $L_{\rm dev}/L_{\rm tot}$ for NDs are well separated from those of the other two, while on average NEs are slightly more bulge-dominated than CQs. Interestingly, the $(c/a)_{\ast}$ distributions of the two bulge-dominated populations are very different: CQs on average are only slightly thicker than NDs but much flatter than NEs. It implies that CQs, although being quenched and bulge-dominated, are still in thin morphologies like NDs, consistent with what we see in Fig~\ref{fig:example_morphology}.

\subsection{Galaxy kinematics}
\label{sec:prop_dynamics}
\begin{figure*}
\includegraphics[width=1.8\columnwidth]{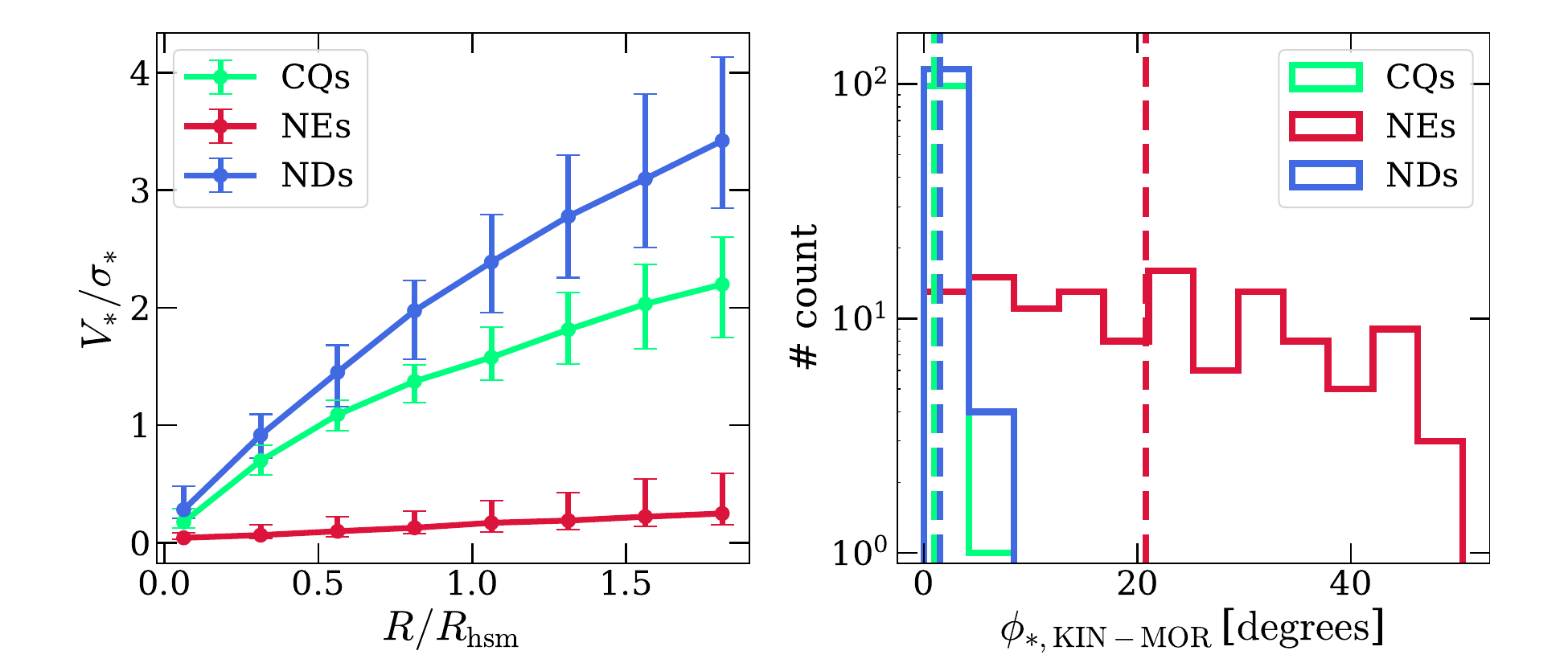}
\caption{Left: the radial $V_{\ast}/\sigma_{\ast}$ profiles for CQs (green), NEs (red), and NDs (blue). The error bars indicate the range from the 16th to the 84th percentiles ($1\sigma$). Right: the distributions of misalignment angles between the photometric and kinematic major axes of stars ($\phi_{\ast,\rm KIN-MOR}$), with the dashed lines indicating the median values of the investigated properties for the three types of galaxies.}
\label{fig:kinematics}
\end{figure*}

To present the differences in galaxy kinematics among the three types of galaxies, we show, in the left panel of Fig.~\ref{fig:kinematics}, the radial profiles of $V_{\ast}/\sigma_{\ast}$ from the edge-on view of the galaxies. For each galaxy, both $V_{\ast}$ and $\sigma_{\ast}$ are calculated in a slit of width 0.5$R_{\rm hsm}$ along the stellar photometric major axis from the edge-on view of the galaxy. The photometric principal axes are obtained by finding eigenvectors of the associated inertia tensors~\citep{Allgood_et_al.(2006)}, using the stellar particles within a sphere with its radius being the smaller one of either $3R_{\rm hsm}$ or 30 kpc, in order to restrict the calculation to the visible galaxy domain. In the right panel, we show the distributions of the misalignment angle $\phi_{\ast,\rm KIN-MOR}$ between the shortest photometric principal axis and the kinematic major axis (the direction of the stellar angular momentum calculated within $2R_{\rm hsm}$). As shown in the figure, NDs exhibit the strongest stellar rotation pattern with $V_{\ast}/\sigma_{\ast}\approx 3.3$ at $r=2R_{\rm hsm}$ and NEs have the weakest rotation with $V_{\ast}/\sigma_{\ast}\approx 0.2$ at $r=2R_{\rm hsm}$. CQs have intermediate rotation with $V_{\ast}/\sigma_{\ast}\approx 2$ at $r=2R_{\rm hsm}$, slightly lower than NDs, but significantly higher than NEs. Interestingly, both NDs and CQs show much better alignments between the kinematic and morphological axes with median values of $\phi_{\ast, \rm KIN-MOR}\sim 1^{\circ}$, in contrast to $\sim 21^{\circ}$ for NEs, whose misalignment angles also span a much broader range up to $\sim50^{\circ}$. Such kinematic properties echo the flatter morphologies (in terms of $(c/a)_{\ast}$) of NDs and CQs in comparison to NEs.

\subsection{Stellar population and star formation}
\label{sec:prop_sp}
\begin{figure}
\includegraphics[width=1\columnwidth]{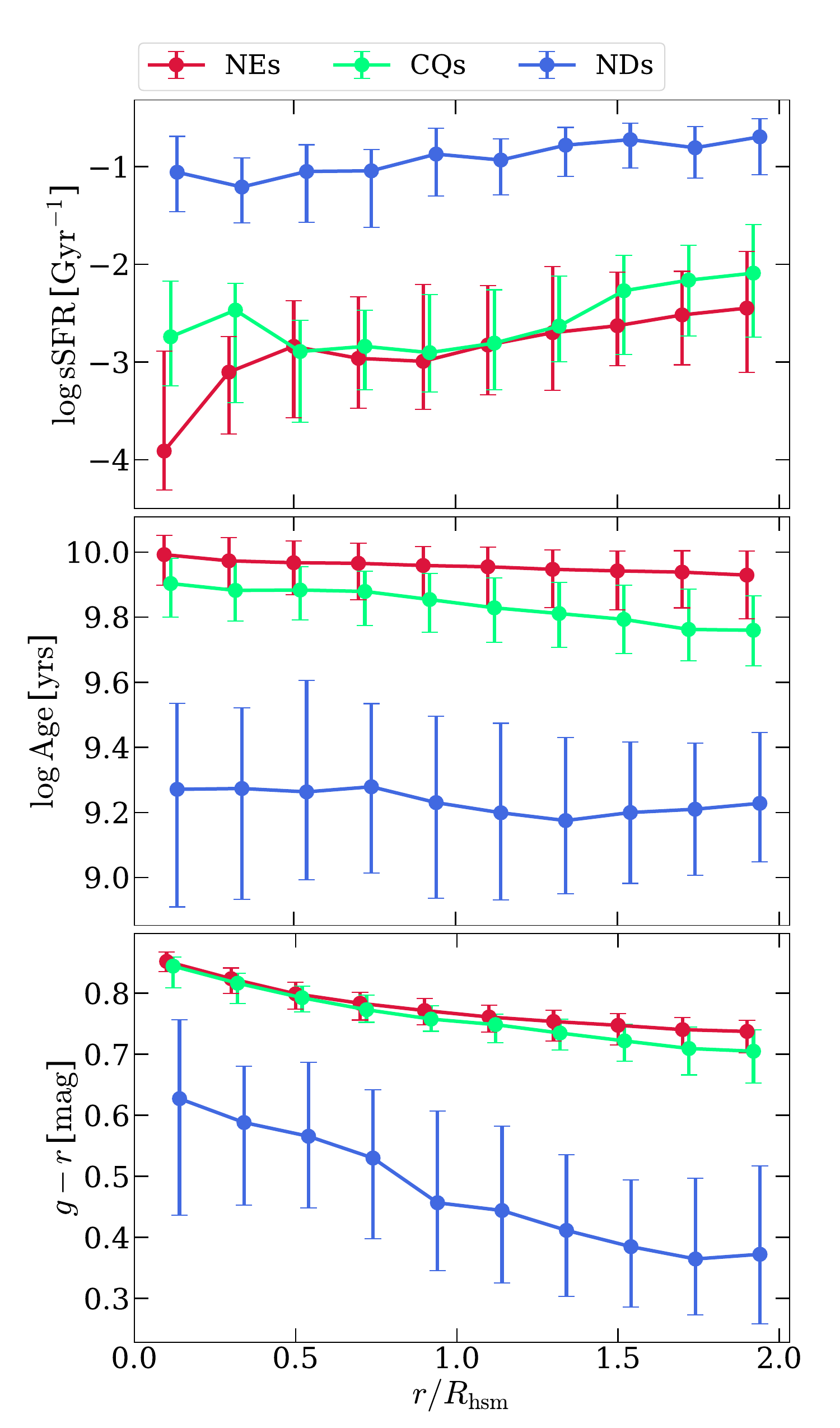}
\caption{The radial profiles of the specific star-formation rate ($\log\mathrm{sSFR}$, top), the stellar age ($\log\mathrm{Age}$, middle), and colour ($g-r$, bottom). CQs, NEs, and NDs are represented by green, red, and blue lines, respectively. The error bars indicate the range from the 16th to the 84th percentiles ($1\sigma$).}
\label{fig:sp}
\end{figure}

\begin{figure*}
\includegraphics[width=1.8\columnwidth]{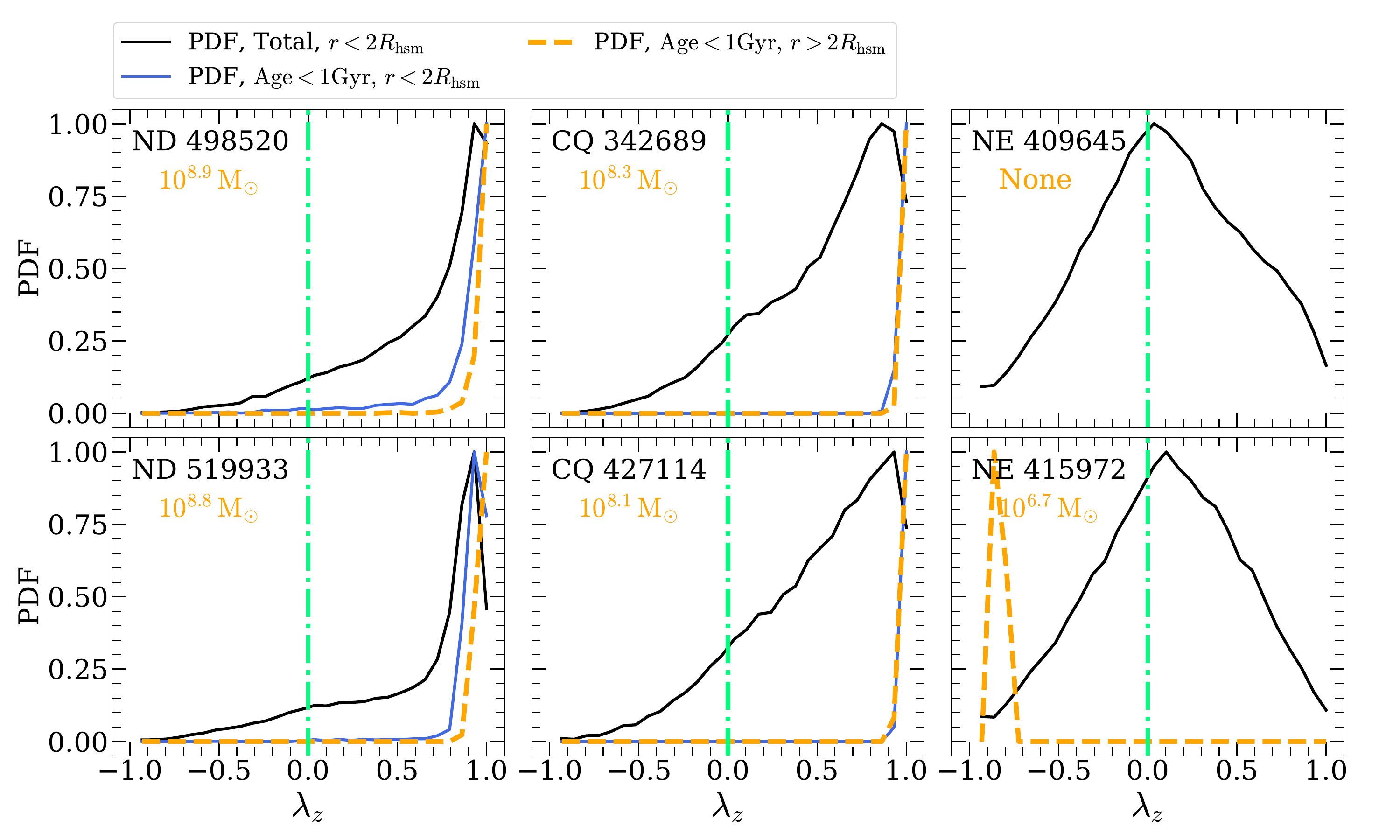}
\caption{The distribution of circularity $\lambda_z$ (see \citealt{Xu_et_al.(2017)} for detailed definition) of two example CQs (ID: 342689, 427114), two example NEs (ID: 409645, 415972), and two example NDs (ID: 498520, 519933) from left to right. 
In each panel, the black solid, the blue solid, and the orange dashed curves represent the particle number probability distribution function (PDF) of all stars within $2R_{\rm hsm}$, the stars formed in the recent $\rm 1\,Gyr$ within $2R_{\rm hsm}$, and the stars formed in the recent $\rm 1\,Gyr$ at the outskirts of the galaxies ($r>2R_{\rm hsm}$), respectively. The value in each panel is the total mass of the newly formed stars (in the recent 1 Gyr) {\it out of} $2R_{\rm hsm}$. The green dotted-dashed line indicates $\lambda_z=0$.}
\label{fig:circularity}
\end{figure*}

In Fig.~\ref{fig:sp}, we show radial profiles of the specific star-formation rate, SDSS $r$-band \citep{Stoughton_et_al.(2002)} luminosity-weighted stellar age, and $g-r$ colour of the three types of galaxies. As can be seen, NDs are much more actively star forming with sSFR $\sim 2$ orders of magnitude higher than that of NEs and CQs. As a result, NDs are significantly younger and bluer than both NEs and CQs. All three types of galaxies exhibit to different extents a decreasing (increasing) trend of sSFR (stellar age and $g-r$ colour) towards galaxy centres, suggesting a gradual ``inside-out'' quenching, as expected (e.g. \citealt{Tacchella_et_al.(2015),Spilker_et_al.(2018),Guo_et_al.(2019),Nelson_et_al.(2021)}). It is worth noting that CQs appear to have more star formation towards galaxy centres and outskirts in comparison to NEs. We further examine the stellar ages among the three types of galaxies. On average, $\sim 40\%$ of the total stellar mass (within $2R_{\rm hsm}$) in a present-day ND was formed in the most recent 5 Gyrs, while the majority of stars in CQs and NEs was formed more than 5 Gyrs ago ($\sim 86\%$ for CQs and $97\%$ for NEs). For the dynamically cold orbits, $\sim 70\%$ of the cold-orbit stars in NDs was formed in the most recent 5 Gyrs, while CQs and NEs had built the majority of their cold orbits more than 5 Gyrs ago ($\sim 70\%$ of CQs and $\sim 90\%$ for NEs), which however, has no longer remained dynamically cold by present time.

In Fig.~\ref{fig:circularity}, we show the distributions of circularity $\lambda_z$ (see \citealt{Xu_et_al.(2017)} for detailed definition) of the total stars within $2R_{\rm hsm}$, the newly formed stars (i.e. formed in the recent 1 Gyr) within $2R_{\rm hsm}$, and the newly formed stars at the outskirts of the galaxies (i.e. $r>2R_{\rm hsm}$) for the six examples (two NDs, two CQs, and two NEs, same as Fig.~\ref{fig:example_morphology}). As expected, the circularity distribution of stars within $2R_{\rm hsm}$ in the dynamically cold galaxies (two NDs and two CQs) peaks at $\lambda_z\sim 1$, while that of NEs peaks at $\lambda_z\sim 0$. Interestingly, the newly formed stars at the outskirts of CQs are non-negligible: the masses of the newly formed stars at the outskirts of CQs are $10^{8.3}\,\mathrm{M_{\odot}}$ and $10^{8.1}\,\mathrm{M_{\odot}}$ for the two CQ examples. The circularity of these newly formed stars also peaks at $\lambda_z\sim 1$, showing the circular orbits of these new stars. This implies that although CQs have ceased their star-formation in the galaxy center (typically within $2R_{\rm hsm}$), new stars are still formed at the outskirts of the galaxies ($r\gtrsim 2R_{\rm hsm}$), which typically inherit the circular motion of the ring-like HI gas (see Fig.~\ref{fig:example_morphology}) and are on cold orbits.

\section{How do they form?}
\label{sec:how_form}
\begin{figure*}
\includegraphics[width=2\columnwidth]{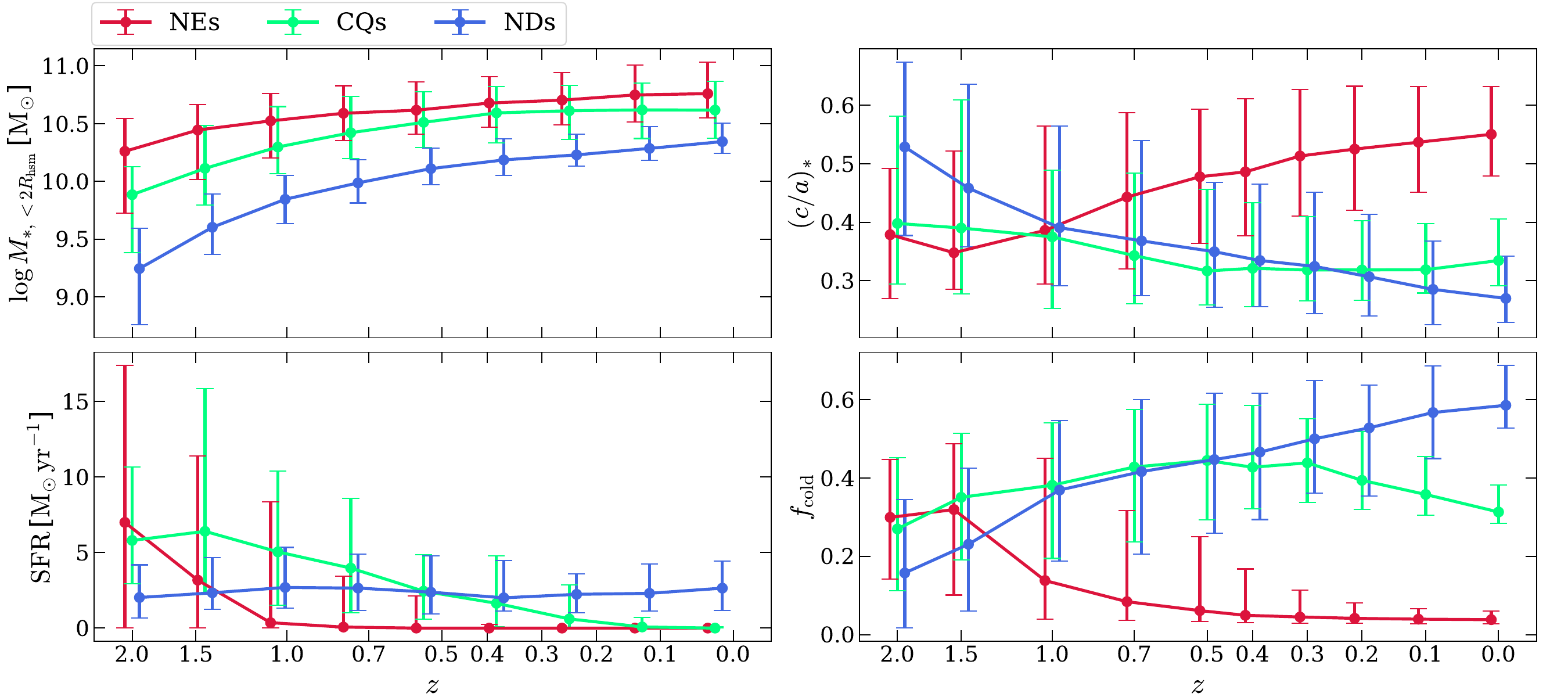}
\caption{Evolution of galaxy properties (top left: the stellar mass within $2R_{\rm hsm}$; top right: the shortest-to-longest principal axis ratio of the stellar component $(c/a)_{\ast}$; bottom left: the star-formation rate (SFR) within $2R_{\rm hsm}$; bottom right: the cold orbit fraction within $2R_{\rm hsm}$) as a function of redshift. In each panel, CQs, NEs, and NDs are indicated by green, red, and blue lines, with error bars indicating the range from the 16th to the 84th percentiles ($1\sigma$).}
\label{fig:paras_evolution}
\end{figure*}

\begin{figure*}
\centering
\includegraphics[width=1.85\columnwidth]{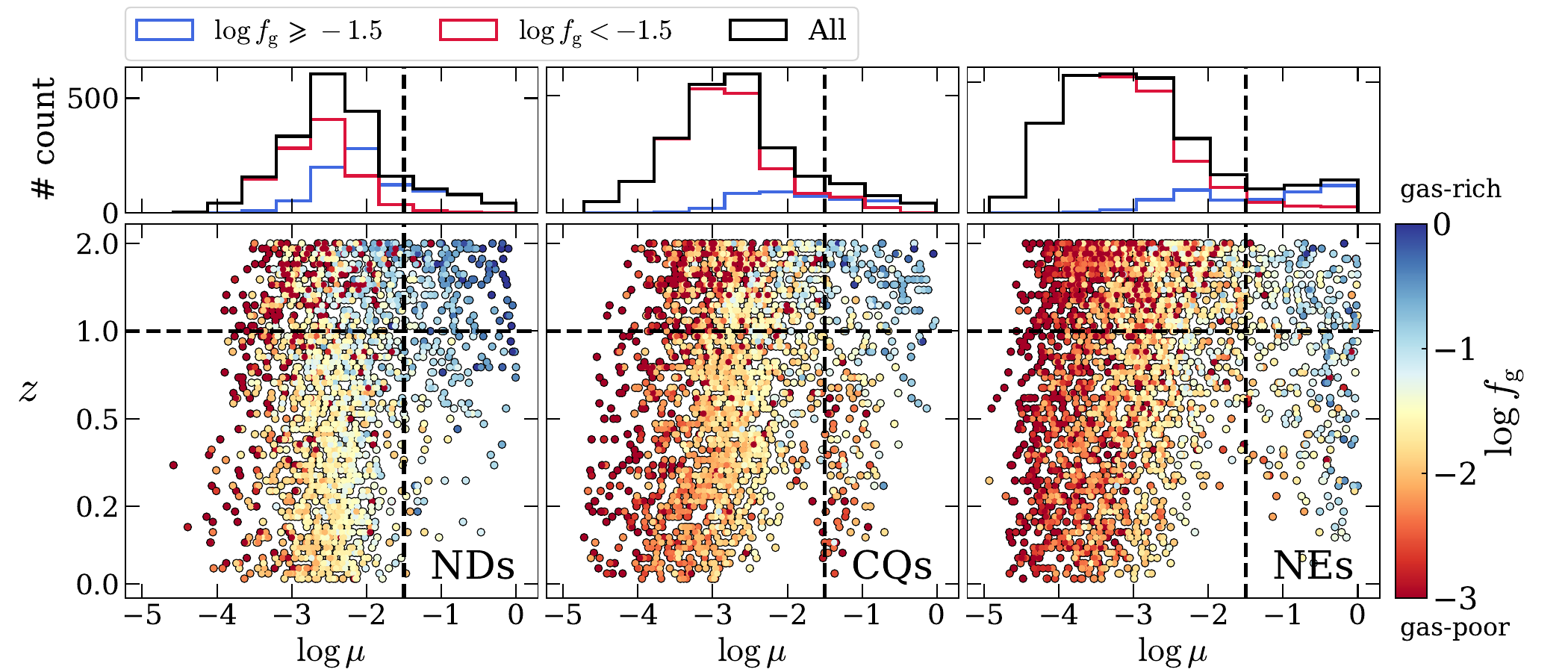}
\includegraphics[width=1.85\columnwidth]{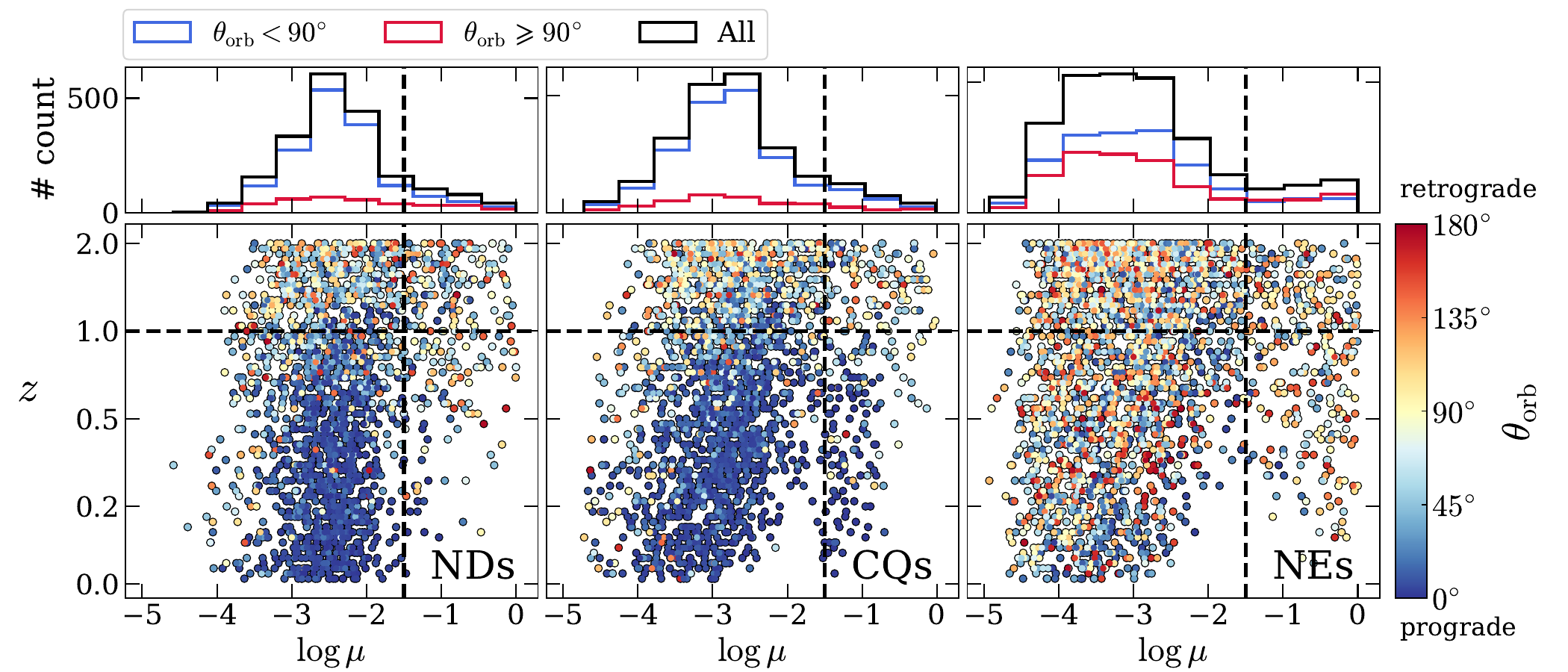}
\includegraphics[width=1.85\columnwidth]{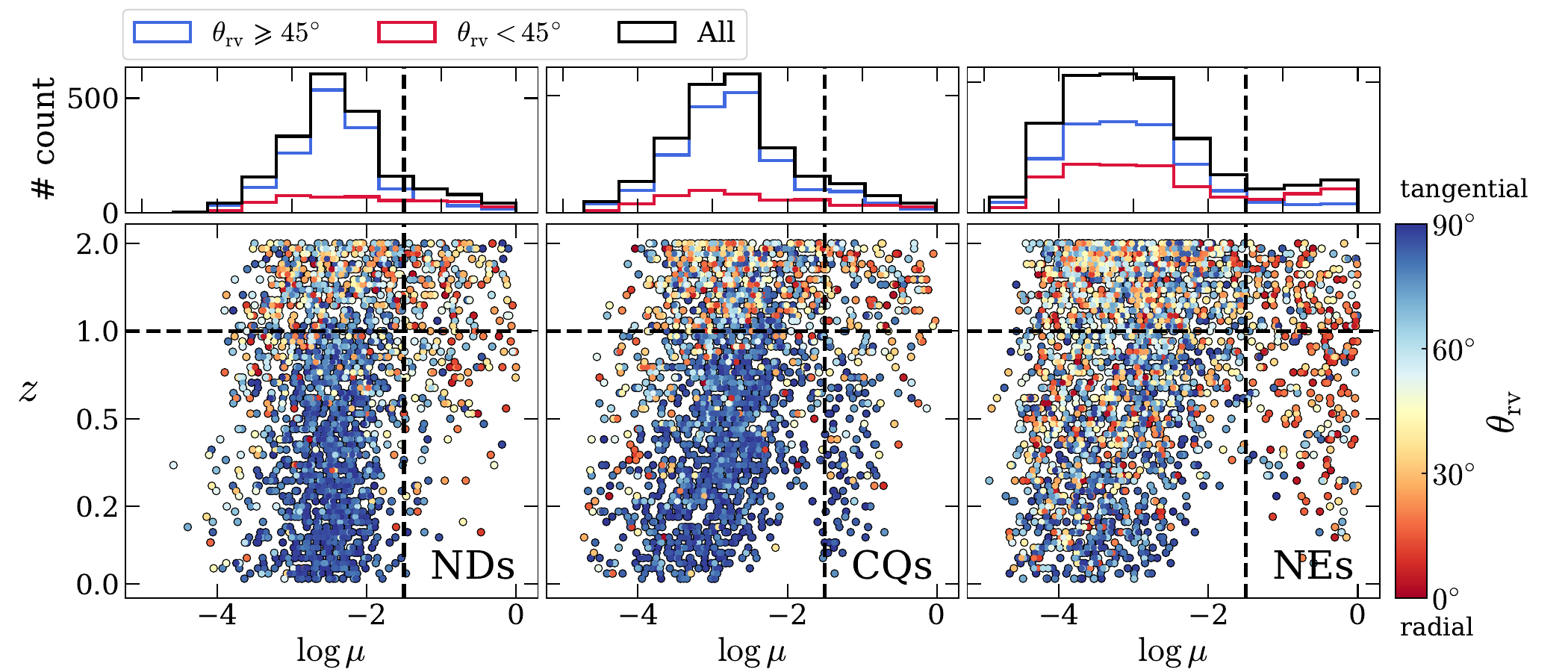}
\caption{Distribution of merger events in the $z-\log\,\mu$ plane ($z$ is the redshift and $\mu$ is the merger stellar mass ratio; see Eq.~\ref{eq:mu} for definition) for NDs (left), CQs (middle), and NEs (right). The colours indicate (1) the merger gas fractions ($f_{\rm g}$), (2) the angle between the host stellar spin axis and the orbital spin axis ($\theta_{\rm orb}$), and (3) the angle between the relative position and the relative velocity of the host and incoming galaxies ($\theta_{\rm rv}$) from top to bottom (see Section~\ref{sec:merger} for definitions). In each panel, each circle represents a merger event that the galaxies experience. The vertical lines indicate $\log\,\mu=-1.5$ (which is the division for major and minor mergers in this work) and the horizontal lines indicate $z=1.0$ (below which CQs and NDs are clearly seen to experience more co-rotating mergers). The histograms on the top of each panel show the distributions of $\log\,\mu$: the black ones indicate the distribution for all mergers; the blue lines indicate results for gas-rich mergers (top panel), prograde mergers (middle panel) and tangential mergers (bottom panel); and the red lines indicate the alternative option in each case.}
\label{fig:merger_info}
\end{figure*}

\begin{figure}
\includegraphics[width=1\columnwidth]{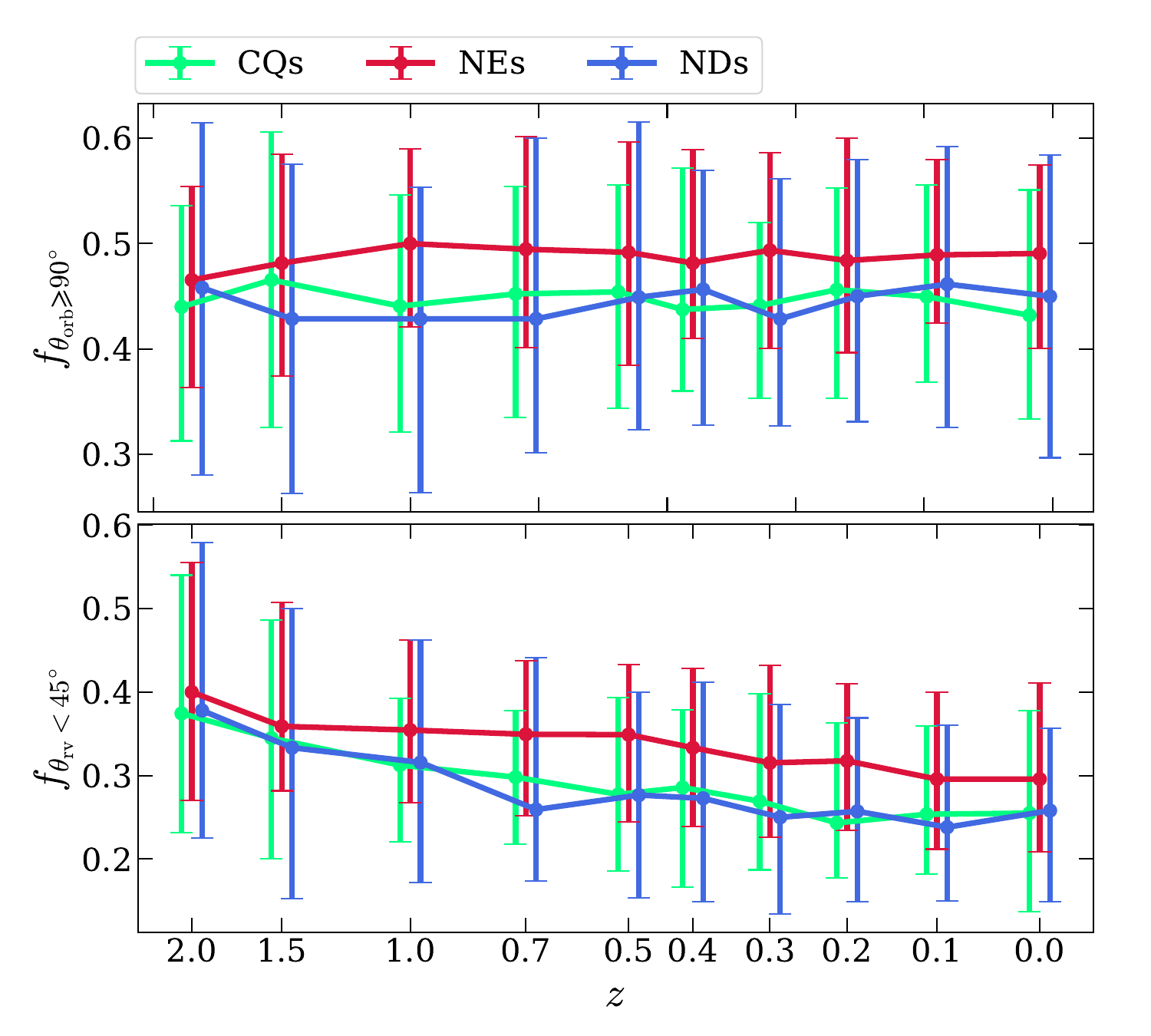}
\caption{The redshift evolution of the number fraction of neighbouring galaxies being retrograde with respect to the stellar spin of the host galaxy (top, $f_{\theta_{\rm orb}\geqslant 90^{\circ}}$) and in radial orbits (bottom, $f_{\theta_{\rm rv}< 45^{\circ}}$). CQs, NEs, and NDs are indicated by green, red, and blue lines, with error bars indicating the range from the 16th to the 84th percentiles ($1\sigma$).}
\label{fig:environment}
\end{figure}
Why do NDs and CQs manage to maintain their cold kinematics while NEs become dynamically hot? To answer this question, we carry out an investigation in Section~\ref{sec:evol_paras} of the redshift evolution of several key parameters of our galaxy samples and in Section~\ref{sec:merger} a statistical analysis of galaxy merging histories.

\subsection{Redshift evolution of general properties}
\label{sec:evol_paras}

In Fig.~\ref{fig:paras_evolution}, we trace the selected galaxies (i.e. NEs, CQs, and NDs) to higher redshifts along their main progenitor branch and present the redshift evolution of their stellar mass within $2R_{\rm hsm}$, $M_{\ast,<2R_{\rm hsm}}$, star-formation rate, SFR, axis ratio, $(c/a)_{\ast}$, and cold-orbit fraction, $f_{\rm cold}$ from $z=2$ to the present. It is worth noting that as demonstrated in \citealt{Rodriguez-Gomez_et_al.(2017)}), galaxies in this mass range have built up the bulk of their stellar mass mainly through in-situ star formation. As can be seen from the figure, NEs and CQs have always been more massive than their ND counterparts, in particular at $z=2$ (the former two are $0.5-1\rm \,dex$ more massive than the latter). Correspondingly the former two have significantly higher sSFR, flatter stellar morphologies, and larger cold-orbit fractions than the latter at $z \geqslant 2$, suggesting an early-phase disc growth and stellar assembly of present-day early-type galaxies, prior to the formation of the present-day star-forming disc galaxies. With time, NDs become flatter and spin up, as indicated by the decrease of $(c/a)_{\ast}$ and the increase of $f_{\rm cold}$ towards lower redshifts (see also \citealt{Pillepich_et_al.(2019)}). CQs follow a similar trend as NDs since $z=1$ and only become slightly thicker and dynamically hotter after $z =0.3 \sim 0.4$, while NEs steadily become thicker and dynamically hotter with decreasing redshift. We note that the SFRs of NEs drop rapidly since $z=2$ and down to quenched states since $z \sim 1$, while CQs continue to build up their star-forming, thin and co-rotating stellar discs until $z = 0.3\sim 0.4$ before being quenched. Interestingly, the SFRs of ND progenitors have never been as high as their early-type counterparts at earlier times, but have remained steady and exhibit constant star formation throughout redshift. This is because the ND progenitors have never had gas reservoirs for star formation as big as their early-type counterparts had to begin with at high redshifts. In fact, if we compare the specific star-formation rates (sSFRs) among the three types of galaxies, CQs and NEs have always had lower sSFRs than their ND counterparts. We also note that some of the ND progenitors at the earliest stages investigated in this work are not well resolved due to their low stellar masses. But we have confirmed that it will not change our statistical results if we exclude these low-mass progenitors.

\subsection{Merging histories establish dynamical status of galaxies}
\label{sec:merger}
In this section, we employ four quantities to describe the merger events the galaxies have experienced: the merger orbit angles $\theta_{\rm orb}$ and $\theta_{\rm rv}$, the merger mass ratio $\mu$, and the merger gas fraction $f_{\rm g}$. All the four properties are measured at the time when the incoming galaxy has the highest stellar mass across its evolution history (see \citealt{Rodriguez-Gomez_et_al.(2015)}). Specifically, $\theta_{\rm orb}$ is defined as:
\begin{equation}
\label{eq:theta}
   \theta_{\rm orb} \equiv | \theta_{\textbf{j}_{\rm orb, incoming}} - \theta_{\textbf{j}_{\ast,\rm host}}|,
\end{equation}
the angle between the orbital angular momentum $\textbf{j}_{\rm orb, incoming}$ of the incoming galaxy (with respect to the center of the host galaxy) and the stellar spin $\textbf{j}_{\ast,\rm host}$ of the host galaxy at the same snapshot. Thus, $\theta_{\rm orb}\geqslant 90^{\circ}$ means that the orbital spin of the incoming galaxy is retrograde with respect to the stellar spin of the host galaxy. $\theta_{\rm rv}$ is defined as:
\begin{equation}
    \theta_{\rm rv} \equiv 90^{\circ} - | \theta_{\Delta \textbf{R}} - \theta_{\Delta \textbf{V}} - 90^{\circ}|,
\end{equation}
where $\Delta \textbf{R}$ and $\Delta \textbf{V}$ are the relative position and velocity of the host and incoming galaxies. Under this definition, $\theta_{\rm rv}\sim 90^{\circ}$ means the merger orbit is tangential and $\theta_{\rm rv}\sim 0^{\circ}$ means radial orbit. The merger mass ratio $\mu$ is defined as:
\begin{equation}
\label{eq:mu}
   \mu \equiv \frac{M_{\ast,\rm incoming}}{M_{\ast,\rm host}},
\end{equation}
where $M_{\ast,\rm incoming}$ and $M_{\ast,\rm host}$ are the stellar masses of the incoming galaxy and the host galaxy within $2R_{\rm hsm}$, respectively. The merger gas fraction is defined as:
\begin{equation}
\label{eq:fg}
    f_{\rm g} \equiv \frac{M_{\rm gas,incoming}}{M_{\ast,\rm incoming}+M_{\ast,\rm host}},
\end{equation}
where $M_{\rm gas,incoming}$ is the gas mass within $2R_{\rm hsm}$ of the incoming galaxy\footnote{Note that the definition of $f_{\rm g}$ here is slightly different from that of \citet{Lu_et_al.(2021a)}.}. In other words, $f_{\rm g}$ traces the gas fraction {\it brought in} by the incoming galaxy to the whole merging system.

In Fig.~\ref{fig:merger_info}, we present, for all three types of galaxies, the key merging properties in the $z-\log\,\mu$ plane, where $z$ is the redshift when a merger occurs. The three panels from upper to bottom are colour-coded by $f_{\rm g}$, $\theta_{\rm orb}$, and $\theta_{\rm rv}$, respectively. A clear bimodal distribution of the merger mass ratio $\mu$ among both CQs and NEs is seen below $z\lesssim 0.7$. At the larger mass-ratio end, NEs have the highest merger mass ratios; CQs come next, while the large-$\mu$ peak for NDs is much less significant. This implies that the large-$\mu$ mergers may be responsible for the formation of the large bulge in CQs and NEs. We note that the distribution of $\mu$ at the smaller mass-ratio end systematically shifts and extends to lower values of $\mu$ towards $z=0$ for CQs and NEs in comparison to NDs, as a result of fixed galaxy resolution and the fact that the former samples are more massive than the latter.  

In terms of gas budgets during mergers, NDs as expected have experienced the most gas-rich mergers across the entire redshift range. It is worth noting that at the large mass-ratio end, NEs (and their progenitors) have experienced many gas-rich large-mass-ratio mergers since $z\sim 0.5$, while CQs (and their progenitors) have experienced the most gas-poor mergers among the three types of galaxies.

From the perspective of merging orbits, CQs and NDs clearly have experienced more frequent low $\theta_{\rm orb}$ (prograde) mergers below $z=1$, while mergers for NEs come with all orbital angles $\theta_{\rm orb}$. The distribution of $\theta_{\rm rv}$ is showing a similar trend as $\theta_{\rm orb}$ on the $z-\log\,\mu$ plane, where the merging orbits that CQs and NDs have experienced are close to tangential orbits (where $\theta_{\rm rv}\sim 90^{\circ}$), while NEs are seen to have more chances to experience radial mergers ($\theta_{\rm rv}\sim 0^{\circ}$). It also implies that most of the prograde merging trajectories (where $\theta_{\rm orb}\sim 0^{\circ}$) that CQs and NDs have experienced are also tangential orbits (where $\theta_{\rm rv}\sim 90^{\circ}$). The difference in the merging orbits reveals a direct reason why NDs and CQs maintain their rotational kinematics and flat morphologies, while NEs become dynamically hot and thicken up. Our findings are also consistent with previous studies, which found that the dynamical friction in massive early-type galaxies may results in radial orbits for major mergers \citep{Pop_et_al.(2018)}, which play a crucial role in the formation of slow-rotating (prolate) galaxies (\citealt{Li_et_al.(2018a)}). We note that we have tried to carry out our statistic analysis above using two different snapshots as the snapshot of merging: (1) the snapshot when the incoming galaxy reached its largest stellar mass historically, after which the stellar mass starts decreasing due to tidal stripping (the currently used); and (2) the snapshot where the incoming galaxy was last identified. We confirmed that our main results keep unchanged under two different definitions.

In Table~\ref{table:table2}, we present the statistics for the key merging properties for the three galaxy types within two different redshift ranges. As can be seen, before $z\sim 1$, on average a present-day NE's progenitor has experienced twice as more mergers as a ND's progenitor, indicating that the former might be born in denser environment than the latter. The latter (ND's progenitor) is seen to have experienced markedly more frequent gas-rich mergers than the former (NE's progenitor); while the former has experienced more frequent low-mass-ratio mergers than the latter. Differences in merging orbits and trajectory types are marginal among the three types of galaxies at these high redshifts. At redshifts below $z \sim 1$, discrepancies among the three types of galaxies become more significant. The progenitors of both present-day NDs and CQs have had dominant fractions of pro-grade and tangentially incoming merging orbits ($\theta_{\rm orb} < 90^{\circ}$ and $\theta_{\rm rv} \geqslant 45^{\circ}$). While a NE's progenitor during that time has experienced significantly more retrograde and radial mergers ($\theta_{\rm orb}\geqslant 90^{\circ}$ and $\theta_{\rm rv} < 45^{\circ}$) than its ND and CQ counterparts. We also note here that some of our quenched early-type galaxies still host plenty of HI gas with high angular momenta at the outskirts of the galaxies (see CQs in Fig.~\ref{fig:example_morphology} for examples). It implies a mechanism which keeps the galaxies quenched and we will address this mechanism in another parallel work \citep{Lu_et_al.(2021b)}.

In Fig.~\ref{fig:environment}, we present the redshift evolution of the number fractions of environment member galaxies which are (1) orbiting in a retrograde fashion (with respect to the stellar spins of the central galaxies, $f_{\theta_{\rm orb}\geqslant 90^{\circ}}$) and (2) orbiting on radial orbits (with respect to the center of the host galaxy, $f_{\theta_{\rm rv}<45^{\circ}}$). Here, we only take the neighbouring galaxies within the same friend-of-friend (FoF, \citealt{Davis_et_al.(1985)}) group, which locate $1R_{\rm htm}$ to $8R_{\rm htm}$ from the center of the investigated central galaxy (where $R_{\rm htm}$ is the half-total-mass radius of the central galaxy) as the environment member galaxies. As can be seen, NEs (and their progenitors) have always had more neighbouring galaxies on retrograde (i.e. $\theta_{\rm orb}\geqslant 90^{\circ}$) and radial (i.e. $\theta_{\rm rv}<45^{\circ}$) orbits than their ND and CQ counterparts, which explains why NEs have experienced more frequent retrograde and radial mergers than the other two types of galaxies. We note that the phenomenon that NEs have higher $f_{\theta_{\rm orb}\geqslant 90^{\circ}}$ is consistent with the fact that the stellar spins of NEs are not well aligned with the environmental (neighbour-galaxy orbital) angular momenta, in comparison to their dynamically cold counterparts. It is worth noting that such a difference in their environment exists as early as $z=2$ when the progenitor of present-day NEs used to be much flatter and have a good sense of rotation allowing for well-defined spin directions. It implies that the present-day dynamically hot ellipticals (NEs) are no longer the evolution remnants of present-day star-forming disc galaxies, but set off on completely different evolution tracks under very different (initial) conditions.

\begin{table*}
\caption{The {\it average} number of mergers within $1<z\,\leqslant\,2$ (top) and below $z=1$ (bottom) per galaxy within different galaxy samples of NDs, CQs, and NEs. See Section~\ref{sec:how_form} for definitions of $f_{\rm g}$, $\mu$, $\theta_{\rm orb}$, and $\theta_{\rm rv}$.}
\setlength{\tabcolsep}{6.5mm}
\begin{tabular}{cccc}
\hline
\hline
 $1<z\,\leqslant\,2$  & NDs & CQs & NEs\\
\hline
\hline
$\log\,\mu\,\geqslant\,-1.5$              & 1.1 (22\%) & 1.2 (15\%) & 1.2 (12\%)\\
$\log\,\mu<-1.5$                          & 3.9 (78\%) & 6.7 (85\%) & 8.8 (88\%)\\
\hline
$\log\,f_{\rm g}\,\geqslant\,-1.5$       & 2.9 (58\%) & 2.5 (32\%) & 1.8 (18\%)\\
$\log\,f_{\rm g}<-1.5$                   & 2.1 (42\%) & 5.4 (68\%) & 8.2 (82\%)\\
\hline
$\theta_{\rm orb}\,\geqslant\,90^{\circ}$ & 1.8 (36\%)  & 2.5 (32\%)  & 4.4 (44\%)\\
$\theta_{\rm orb} <90^{\circ}$            & 3.2 (64\%)  & 5.4 (68\%)  & 5.6 (56\%)\\
\hline
$\theta_{\rm rv}\,\geqslant\,45^{\circ}$  & 2.8 (56\%)  & 4.6 (58\%)  & 5.4 (54\%)\\
$\theta_{\rm rv} <45^{\circ}$             & 2.2 (44\%)  & 3.3 (42\%)  & 4.6 (46\%)\\
\hline

Total                                     & 5.0 & 7.9 & 10.0\\
\hline
\hline
\end{tabular}

\begin{tabular}{cccc}
\hline
\hline
$z \leqslant 1$    & NDs & CQs & NEs\\
\hline
\hline
$\log\,\mu\,\geqslant\,-1.5$              & 0.9 (8\%) & 1.3 (9\%) & 1.5 (12\%)\\
$\log\,\mu<-1.5$                          & 10.0 (92\%) & 12.8 (91\%) & 11.0 (88\%)\\
\hline
$\log\,f_{\rm g}\,\geqslant\,-1.5$       & 4.2 (38\%) & 1.4 (10\%) & 1.8 (14\%)\\
$\log\,f_{\rm g}<-1.5$                   & 6.7 (62\%) & 12.7 (90\%) & 10.7 (86\%)\\
\hline
$\theta_{\rm orb}\,\geqslant\,90^{\circ}$ & 1.1 (9\%)  & 1 (7\%)  & 5.0 (40\%)\\
$\theta_{\rm orb} <90^{\circ}$            & 9.8 (91\%) & 13.1 (93\%) & 7.5 (60\%)\\
\hline
$\theta_{\rm rv}\,\geqslant\,45^{\circ}$  & 9.5 (87\%) & 12.7 (90\%) & 8.2 (66\%)\\
$\theta_{\rm rv} <45^{\circ}$             & 1.4 (13\%)  & 1.4 (10\%)  & 4.3 (34\%)\\
\hline

Total                                     & 10.9 & 14.1 & 12.5\\
\hline
\hline
\end{tabular}

\vspace{2mm}
\label{table:table2}
\end{table*}

\section{Real-world Counterparts and Observational Predictions}
\label{sec:observation}
\begin{figure}
\includegraphics[width=1.0\columnwidth]{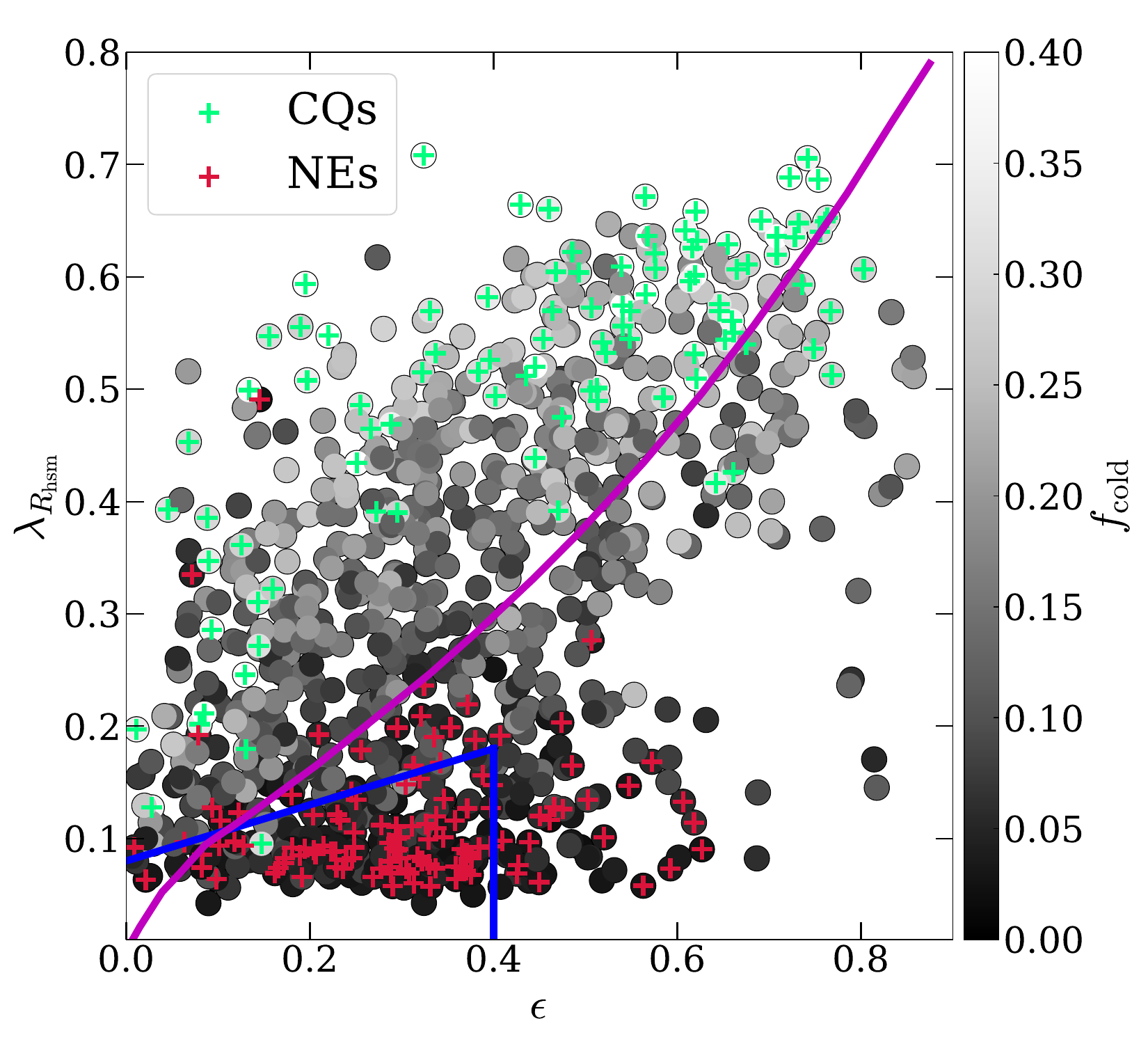}
\caption{Distributions of early-type galaxies (see Section~\ref{sec:sample} for definition) in the $\lambda_{R_{\rm hsm}}-\epsilon$ plane ,where $\lambda_{R_{\rm hsm}}$ is the stellar spin parameter (measured within $r=R_{\rm hsm}$) and $\epsilon$ (the ellipticity measured at $r=3R_{\rm hsm}$). The colours indicate the cold orbital fraction of the galaxies. The selected CQs and NEs are indicated with green and red plus symbols, respectively. The blue lines represent $\lambda_{R_{\rm hsm}}=\epsilon/4+0.08$ and $\epsilon=0.4$, within (beyond) which galaxies are defined to be slow (fast) rotators. The magenta curve shows the edge-on view for ellipsoidal galaxies integrated up to infinity with $\beta=0.7\epsilon$ (where $\beta$ is the anisotropy parameter, see more details in \citealt{Emsellem_et_al.(2011)}).}
\label{fig:rotators}
\end{figure}

More and more studies of integral-field data have pointed out that early-type galaxies consist of two distinct populations based on their kinematics: fast and slow rotators (e.g. \citealt{Cappellari_et_al.(2006),Cappellari_et_al.(2007),Cappellari_et_al.(2013),Emsellem_et_al.(2007),Emsellem_et_al.(2011)}; see \citealt{Cappellari(2016)} for a review), with galaxies in the former class exhibiting a clear large-scale rotation structure and those in the latter showing no significant rotation. This is naturally linked to our CQ and NE galaxies. We present Fig.~\ref{fig:rotators}, where we show the distributions of our selected early-type galaxy samples in the $\lambda_{R_{\rm hsm}}-\epsilon$ plane. To calculate $\lambda_{R_{\rm hsm}}$ and $\epsilon$, we first project our galaxies along the $Z-$axis of the simulation box and divide the galaxy image into pixels. The ellipticity $\epsilon$ is then calculated as $1-b/a$, where $b/a$ is the minor-to-major axis ratio of galaxies from the $Z-$projection calculated within $3R_{\rm hsm}$. The dimensionless spin parameter $\lambda_{R_{\rm hsm}}$ is the stellar spin parameter within $R_{\rm hsm}$, which is defined as:
\begin{equation}
    \lambda_{R_{\rm hsm}} = \frac{\sum_{i} L_i R_i |V_i|}{\sum_{i} L_i R_i \sqrt{V_i^2+\sigma_i^2}},
\end{equation}
where $L_i$ is the luminosity of the $i-$th pixel; $R_i$ is the distance of the $i-$th pixel to the galaxy center; $V_i$ and $\sigma_i$ are the mean line-of-sight velocity and dispersion of the $i-$th pixel. The summation goes over all the pixels within $R_{\rm hsm}$. We note that the aperture size (the stellar half mass radius in the 3D space) used here is different from what was used in previous studies (i.e. the half-light isophote; see \citealt{Emsellem_et_al.(2007)}). However, in this work, using $R_{\rm hsm}$ is already good enough for a rough classification of fast and slow rotators. As can be seen in the figure, most NEs lie in the region of slow rotators while CQs are seen to be located in the region of extreme fast rotators (higher $\lambda_{R_{\rm hsm}}$ at the same $\epsilon$). \citet{Cappellari(2016)} also showed that most S0 galaxies occupy the same region as fast rotators (see fig. 15 in \citealt{Cappellari(2016)}). Hence, we propose that the observed fast-rotating early-type galaxies and some S0 galaxies are the real-word counterparts of our dynamically cold quenched early-type galaxies (CQs) and the observed slow rotators are the real-word counterparts of the dynamically hot ellipticals (NEs) in this study. We further point out that our study have revealed the origins of the two distinct populations of early-type galaxies (i.e. slow and fast rotators): slow rotators have experienced more retrograde and radial mergers, while fast rotators have experienced more prograde and tangential mergers. The difference in their merger histories may stem from the difference in their environments (see Fig.~\ref{fig:environment}).

We also note that a series of recent studies on observed red spiral galaxies found that these galaxies possess compact cores in their centres and very often exhibit bar and ring/shell features further out \citep{Guo_et_al.(2020)}. In terms of their present-day stellar populations, kinematics (see fig. 2 of \citealt{Hao_et_al.(2019)}, in comparison to Fig.~\ref{fig:kinematics} in this study), and star formation histories \citep{Zhou_et_al.(2020)}, they are more similar to ellipticals than to blue spiral galaxies. The bulk of their stellar mass was found being built up before $z\sim 2$ through a fast formation process. More importantly, these galaxies are not likely to be progenitors of present-day blue spiral galaxies, but are fundamentally different from an early-phase onwards \citep{Zhou_et_al.(2020)}. Indeed a good fraction of the red spiral galaxies in these studies are also classified as S0-type galaxies. We would like to address that the investigations on CQs in our work suggest many similarities to the observations above of red spiral galaxies, even though the two populations are selected using different specific criteria.  

\begin{figure*}
\includegraphics[width=1\columnwidth]{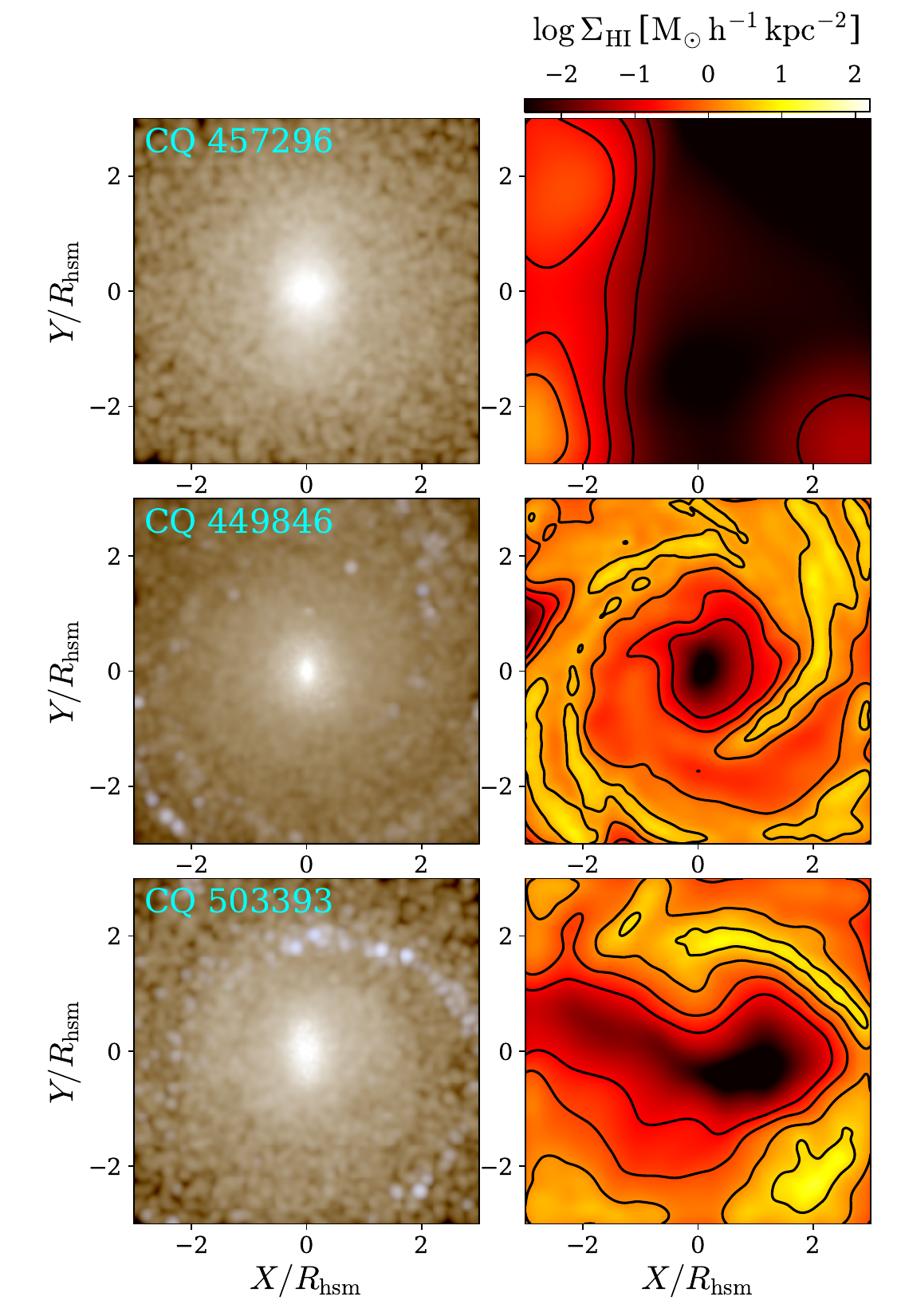}
\includegraphics[width=1\columnwidth]{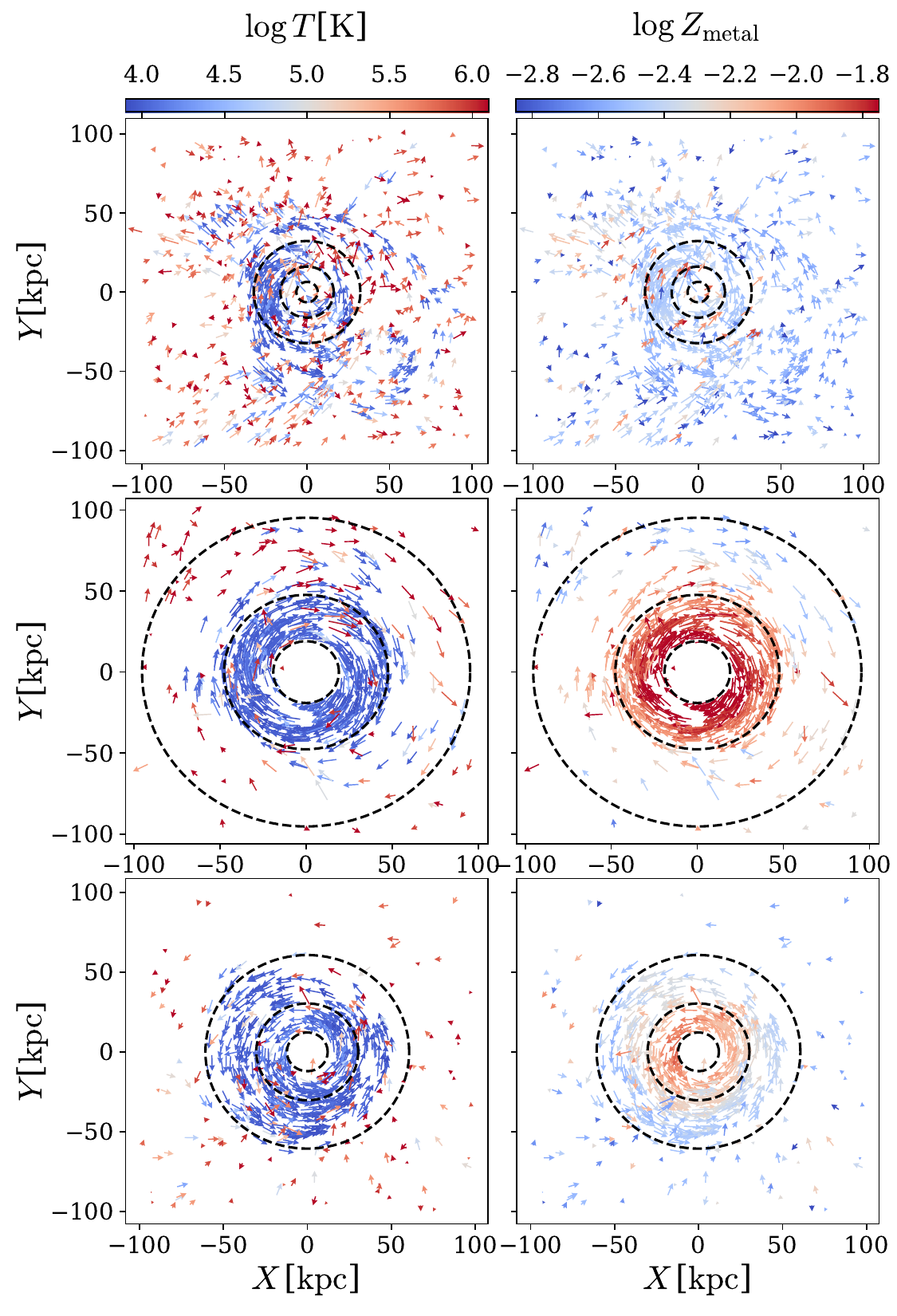}
\caption{Three further examples of CQs (IDs: 457296, 449846, and 503393 from top to bottom, with $\log\,M_{\ast}/\mathrm{M_{\odot}}$ being 10.5, 10.9, and 10.6, respectively). For each example, from left to right, we show: (1) the RGB image, (2) the surface density map of HI gas, (3) the velocity field of outer gas ($2R_{\rm hsm}<r<100\,\mathrm{kpc}$), colour-coded by the gas temperature, and (4) the velocity field of outer gas, colour-coded by the gas metallicity. In the last two columns, the arrows indicate the velocity vectors of the gas cells on the $X-Y$ plane, with their lengths denoting the magnitude of the velocities. Gas cells are plotted every 40 of them. The black dashed circles indicate $r=2R_{\rm hsm},\,5R_{\rm hsm}$, and $10R_{\rm hsm}$ from inner to outer. All these galaxies are viewed from their face-on views.}
\label{fig:ce_rgb}
\end{figure*}

From the view point of the HI gas environment, \citet{Zhang_et_al.(2019)} reported detections of significant HI gas reservoirs around quenched disc galaxies. \citet{Cortese_et_al.(2020)}, however, argued otherwise that the statistical results from \citet{Zhang_et_al.(2019)} were subject to the adopted definition for star formation rate. We note that the questioned samples, in which HI gas reservoirs were detected, indeed exhibit inner passive bulges and outer star-forming discs, as seen in the SDSS images (see fig. 3 in \citealt{Cortese_et_al.(2020)}). These galaxies bear a resemblance to our CQ sample in this study. In Fig.~\ref{fig:ce_rgb}, we show three further example CQ galaxies in terms of their RGB images and local HI gas morphologies in order to provide some hints for the  HI ring detections in observations. We note that the majority of the CQs have optical morphologies of classical elliptical galaxies with significantly low central\footnote{We note that the ``central'' parts of galaxies in this paper refer to the regions where $r<2R_{\rm hsm}$.} gas content (like the one shown in the top panel). However, roughly 40\% of the CQs (like the two examples shown in the two panels below) also exhibit star-forming (spiral-shaped) rings/shells around $r\gtrsim 2R_{\rm hsm}$. As can be seen from the right two columns of Fig.~\ref{fig:ce_rgb}, the surrounding CGM gas show co-rotating kinematics, with temperature in cold (or warm) phase and metallicity being either metal-rich or metal-poor. We therefore make specific predictions that spatially resolved radio observations may reveal a good fraction ($\sim 40\%$) of fast-rotating early-type/S0-type galaxies exhibiting ring-like HI gas structures with high angular momenta at the outskirts of the optical galaxies. Besides, absorption line observations along background quasar sight lines (see \citealt{Stewart(2017),Tumlinson_et_al.(2017)} and the references therein) may also be used to probe the CGM spins around the fast-rotating early-type galaxies/S0s.

\section{Conclusions and discussion}
\label{sec:conclusion_and_discussion}
Quenched early-type (bulge-dominated) galaxies are commonly associated with hot kinematics. However both observations (e.g., \citealt{Emsellem_et_al.(2007)}) and cosmological simulations (e.g., \citealt{Xu_et_al.(2019)}) have reported the presence of quenched but dynamically cold early-type galaxies. In this work, we make use of an advanced cosmological simulation -- the IllustrisTNG Simulation -- to study the properties and origin of the dynamically cold quenched bulge-dominated galaxies (CQs) at the more massive end of the mass spectrum. In particular, through comparisons to a sample of dynamically cold star-forming disc galaxies (NDs) and a sample of dynamically hot quenched elliptical galaxies (NEs), we attempt to address the question regarding different formation and evolution scenarios of three different types of galaxies. The main conclusions are given below, followed by more discussions.

\begin{enumerate}
    \item By sample construction, both CQs and NEs are required to have bulge-to-toal ratio $L_{\rm dev}/L_{\rm tot}>0.5$. However, CQs are as flat as NDs in terms of the axis ratio $(c/a)_{\ast}$ (Fig.~\ref{fig:morphology}). CQs and NDs have similar $V_{\ast}/\sigma_{\ast}$ radial profiles, which indicates significant rotation of the stellar components. Both types of galaxies also have well-aligned photometric and kinematic axes. In contrast, NEs have very small $V_{\ast}/\sigma_{\ast}$ and exhibit markedly large kinematic misalignment angles (Fig.~\ref{fig:kinematics}). However the stellar populations in both quenched CQs and NEs are much older in age and redder in colour, compared to their star-forming disc counterparts (Fig.~\ref{fig:sp}). Besides, CQs are found to form stars (although at a low level) at the outskirts of the galaxies, which inherit the circular motion of the ring-like HI gas (Fig.~\ref{fig:circularity}).

    \item At high redshift ($z\sim 2$), the progenitors of the present-day quenched early-type galaxies (i.e. CQs and NEs) are more massive, more actively star-forming, flatter, and more rotational than their ND progenitor counterparts, indicating the former two went through an early-phase fast disc assembly compared to the present-day disc galaxies. Since then, NEs begin to heat up with their star-formation ceasing quickly, while CQs and NDs continue to build up their stellar disc (decreasing $(c/a)_{\ast}$ and increasing $f_{\rm cold}$) until $z=0.3\sim 0.4$, after which CQs gradually run out of their central ``fuel'' for star formation and become quenched at present times (Fig.~\ref{fig:paras_evolution}).

    \item The three types of galaxies are found to have experienced very different merging histories. More specifically, NEs have experienced more frequent retrograde and radial mergers across the whole redshift range, while CQs and NDs have experienced much more frequent prograde and tangential mergers since $z \sim 1$. This explains the much higher fractions of dynamically cold orbits in CQs and NDs than in NEs. We also note that CQs and NEs have experienced more larger mass-ratio mergers below $z\sim 0.5$ than their ND counterparts (Fig.~\ref{fig:merger_info}). These differences in mergers can be explained by the difference in their environments (Fig.~\ref{fig:environment}).
    
    \item Observed fast and slow rotating early-type galaxies are likely to be the real-world counterparts of our selected CQs and NEs (Fig.~\ref{fig:rotators}). We propose that spatially resolved radio observations may reveal ring-like HI gas structures with high angular momenta at the outskirts of fast rotators and some S0 galaxies (Fig.~\ref{fig:ce_rgb}).
\end{enumerate}

As mentioned in Section~\ref{sec:how_form}, NEs, although are quenched, have experienced many gas-rich large-$\mu$ mergers, even at low redshifts. It offers a challenge to explain why they keep the quenching status under such gas-rich merging histories. Figs.~\ref{fig:example_morphology} and \ref{fig:ce_rgb} have shown that many of the quenched early-type galaxies still maintain high-angular-momenta CGM gas outside the galaxy centres (i.e. $r\gtrsim 2R_{\rm hsm}$). Thus, it is natural to ask: is there any connection between the quenching status in these galaxies and the high-angular-momenta CGM gas at the outskirts of the galaxies? And further, what causes the high CGM angular momentum in these galaxies? We address these interesting questions in our another series of studies \citep{Wang_et_al.(2021),Lu_et_al.(2021b)}, in which the connections between the environment torque field, the CGM gas, and the inner star-formation activities in galaxies are studied.


\section*{Acknowledgements}
We would like to thank Drs. Volker Springel, Zheng Cai, Cheng Li, Caina Hao, Jing Wang, Lan Wang, Yougang Wang, Shihong Liao, Rui Guo, and Mr. Guangquan Zeng for constructive and insightful suggestions and comments which improved the paper. This work is partly supported by the National Key Research and Development Program of China (No. 2018YFA0404501 to SM), by the National Science Foundation of China (Grant No. 11821303, 11761131004 and 11761141012). DX also thanks the Tsinghua University Initiative Scientific Research Program ID 2019Z07L02017. YW acknowledges the support of a Stanford-KIPAC Chabolla Fellowship. 
\section*{Data availability}
General properties of the galaxies in the IllustrisTNG Simulation is available from \url{http://www.tng-project.org/data/}. The rest of the data underlying the article will be shared on reasonable request to the corresponding author.

\bibliographystyle{mnras}
\bibliography{ref}

\begin{thebibliography}{}
\makeatletter
\relax
\def\mn@urlcharsother{\let\do\@makeother \do\$\do\&\do\#\do\^\do\_\do\%\do\~}
\def\mn@doi{\begingroup\mn@urlcharsother \@ifnextchar [ {\mn@doi@}
  {\mn@doi@[]}}
\def\mn@doi@[#1]#2{\def\@tempa{#1}\ifx\@tempa\@empty \href
  {http://dx.doi.org/#2} {doi:#2}\else \href {http://dx.doi.org/#2} {#1}\fi
  \endgroup}
\def\mn@eprint#1#2{\mn@eprint@#1:#2::\@nil}
\def\mn@eprint@arXiv#1{\href {http://arxiv.org/abs/#1} {{\tt arXiv:#1}}}
\def\mn@eprint@dblp#1{\href {http://dblp.uni-trier.de/rec/bibtex/#1.xml}
  {dblp:#1}}
\def\mn@eprint@#1:#2:#3:#4\@nil{\def\@tempa {#1}\def\@tempb {#2}\def\@tempc
  {#3}\ifx \@tempc \@empty \let \@tempc \@tempb \let \@tempb \@tempa \fi \ifx
  \@tempb \@empty \def\@tempb {arXiv}\fi \@ifundefined
  {mn@eprint@\@tempb}{\@tempb:\@tempc}{\expandafter \expandafter \csname
  mn@eprint@\@tempb\endcsname \expandafter{\@tempc}}}

\bibitem[\protect\citeauthoryear{{Allgood}, {Flores}, {Primack}, {Kravtsov},
  {Wechsler}, {Faltenbacher}  \& {Bullock}}{{Allgood}
  et~al.}{2006}]{Allgood_et_al.(2006)}
{Allgood} B.,  {Flores} R.~A.,  {Primack} J.~R.,  {Kravtsov} A.~V.,  {Wechsler}
  R.~H.,  {Faltenbacher} A.,   {Bullock} J.~S.,  2006, \mn@doi [\mnras]
  {10.1111/j.1365-2966.2006.10094.x}, \href
  {https://ui.adsabs.harvard.edu/abs/2006MNRAS.367.1781A} {367, 1781}

\bibitem[\protect\citeauthoryear{{Barnes}}{{Barnes}}{1992}]{Barnes_et_al.(1992)}
{Barnes} J.~E.,  1992, \mn@doi [\apj] {10.1086/171522}, \href
  {https://ui.adsabs.harvard.edu/abs/1992ApJ...393..484B} {393, 484}

\bibitem[\protect\citeauthoryear{{Barnes}}{{Barnes}}{2002}]{Barnes_et_al.(2002)}
{Barnes} J.~E.,  2002, \mn@doi [\mnras] {10.1046/j.1365-8711.2002.05335.x},
  \href {https://ui.adsabs.harvard.edu/abs/2002MNRAS.333..481B} {333, 481}

\bibitem[\protect\citeauthoryear{{Barnes} \& {Hernquist}}{{Barnes} \&
  {Hernquist}}{1991}]{Barnes_et_al.(1991)}
{Barnes} J.~E.,  {Hernquist} L.~E.,  1991, \mn@doi [\apjl] {10.1086/185978},
  \href {https://ui.adsabs.harvard.edu/abs/1991ApJ...370L..65B} {370, L65}

\bibitem[\protect\citeauthoryear{{Barnes} \& {Hernquist}}{{Barnes} \&
  {Hernquist}}{1996}]{Barnes_et_al.(1996)}
{Barnes} J.~E.,  {Hernquist} L.,  1996, \mn@doi [\apj] {10.1086/177957}, \href
  {https://ui.adsabs.harvard.edu/abs/1996ApJ...471..115B} {471, 115}

\bibitem[\protect\citeauthoryear{{Bois} et~al.,}{{Bois}
  et~al.}{2011}]{Bois_et_al.(2011)}
{Bois} M.,  et~al., 2011, \mn@doi [\mnras] {10.1111/j.1365-2966.2011.19113.x},
  \href {https://ui.adsabs.harvard.edu/abs/2011MNRAS.416.1654B} {416, 1654}

\bibitem[\protect\citeauthoryear{{Bournaud}, {Jog}  \& {Combes}}{{Bournaud}
  et~al.}{2005}]{Bournaud_et_al.(2005)}
{Bournaud} F.,  {Jog} C.~J.,   {Combes} F.,  2005, \mn@doi [\aap]
  {10.1051/0004-6361:20042036}, \href
  {https://ui.adsabs.harvard.edu/abs/2005A&A...437...69B} {437, 69}

\bibitem[\protect\citeauthoryear{{Bournaud}, {Jog}  \& {Combes}}{{Bournaud}
  et~al.}{2007}]{Bournaud_et_al.(2007)}
{Bournaud} F.,  {Jog} C.~J.,   {Combes} F.,  2007, \mn@doi [\aap]
  {10.1051/0004-6361:20078010}, \href
  {https://ui.adsabs.harvard.edu/abs/2007A&A...476.1179B} {476, 1179}

\bibitem[\protect\citeauthoryear{{Cappellari}}{{Cappellari}}{2016}]{Cappellari(2016)}
{Cappellari} M.,  2016, \mn@doi [\araa] {10.1146/annurev-astro-082214-122432},
  \href {https://ui.adsabs.harvard.edu/abs/2016ARA&A..54..597C} {54, 597}

\bibitem[\protect\citeauthoryear{{Cappellari} et~al.,}{{Cappellari}
  et~al.}{2006}]{Cappellari_et_al.(2006)}
{Cappellari} M.,  et~al., 2006, \mn@doi [\mnras]
  {10.1111/j.1365-2966.2005.09981.x}, \href
  {https://ui.adsabs.harvard.edu/abs/2006MNRAS.366.1126C} {366, 1126}

\bibitem[\protect\citeauthoryear{{Cappellari} et~al.,}{{Cappellari}
  et~al.}{2007}]{Cappellari_et_al.(2007)}
{Cappellari} M.,  et~al., 2007, \mn@doi [\mnras]
  {10.1111/j.1365-2966.2007.11963.x}, \href
  {https://ui.adsabs.harvard.edu/abs/2007MNRAS.379..418C} {379, 418}

\bibitem[\protect\citeauthoryear{{Cappellari} et~al.,}{{Cappellari}
  et~al.}{2013}]{Cappellari_et_al.(2013)}
{Cappellari} M.,  et~al., 2013, \mn@doi [\mnras] {10.1093/mnras/stt644}, \href
  {https://ui.adsabs.harvard.edu/abs/2013MNRAS.432.1862C} {432, 1862}

\bibitem[\protect\citeauthoryear{{Cortese}, {Catinella}, {Cook}  \&
  {Janowiecki}}{{Cortese} et~al.}{2020}]{Cortese_et_al.(2020)}
{Cortese} L.,  {Catinella} B.,  {Cook} R.~H.~W.,   {Janowiecki} S.,  2020,
  \mn@doi [\mnras] {10.1093/mnrasl/slaa032}, \href
  {https://ui.adsabs.harvard.edu/abs/2020MNRAS.494L..42C} {494, L42}

\bibitem[\protect\citeauthoryear{{Davis}, {Efstathiou}, {Frenk}  \&
  {White}}{{Davis} et~al.}{1985}]{Davis_et_al.(1985)}
{Davis} M.,  {Efstathiou} G.,  {Frenk} C.~S.,   {White} S.~D.~M.,  1985,
  \mn@doi [\apj] {10.1086/163168}, \href
  {https://ui.adsabs.harvard.edu/abs/1985ApJ...292..371D} {292, 371}

\bibitem[\protect\citeauthoryear{{Davis} et~al.,}{{Davis}
  et~al.}{2011}]{Davis_et_al.(2011)}
{Davis} T.~A.,  et~al., 2011, \mn@doi [\mnras]
  {10.1111/j.1365-2966.2011.19355.x}, \href
  {https://ui.adsabs.harvard.edu/abs/2011MNRAS.417..882D} {417, 882}

\bibitem[\protect\citeauthoryear{{Dolag}, {Borgani}, {Murante}  \&
  {Springel}}{{Dolag} et~al.}{2009}]{Dolag_et_al.(2009)}
{Dolag} K.,  {Borgani} S.,  {Murante} G.,   {Springel} V.,  2009, \mn@doi
  [\mnras] {10.1111/j.1365-2966.2009.15034.x}, \href
  {http://adsabs.harvard.edu/abs/2009MNRAS.399..497D} {399, 497}

\bibitem[\protect\citeauthoryear{{Emsellem} et~al.,}{{Emsellem}
  et~al.}{2007}]{Emsellem_et_al.(2007)}
{Emsellem} E.,  et~al., 2007, \mn@doi [\mnras]
  {10.1111/j.1365-2966.2007.11752.x}, \href
  {https://ui.adsabs.harvard.edu/abs/2007MNRAS.379..401E} {379, 401}

\bibitem[\protect\citeauthoryear{{Emsellem} et~al.,}{{Emsellem}
  et~al.}{2011}]{Emsellem_et_al.(2011)}
{Emsellem} E.,  et~al., 2011, \mn@doi [\mnras]
  {10.1111/j.1365-2966.2011.18496.x}, \href
  {https://ui.adsabs.harvard.edu/abs/2011MNRAS.414..888E} {414, 888}

\bibitem[\protect\citeauthoryear{{Genel} et~al.,}{{Genel}
  et~al.}{2014}]{Genel_et_al.(2014)}
{Genel} S.,  et~al., 2014, \mn@doi [\mnras] {10.1093/mnras/stu1654}, \href
  {http://adsabs.harvard.edu/abs/2014MNRAS.445..175G} {445, 175}

\bibitem[\protect\citeauthoryear{{Guo} et~al.,}{{Guo}
  et~al.}{2019}]{Guo_et_al.(2019)}
{Guo} K.,  et~al., 2019, \mn@doi [\apj] {10.3847/1538-4357/aaee88}, \href
  {https://ui.adsabs.harvard.edu/abs/2019ApJ...870...19G} {870, 19}

\bibitem[\protect\citeauthoryear{{Guo}, {Hao}, {Xia}, {Shi}, {Chen}, {Li}  \&
  {Gu}}{{Guo} et~al.}{2020}]{Guo_et_al.(2020)}
{Guo} R.,  {Hao} C.-N.,  {Xia} X.,  {Shi} Y.,  {Chen} Y.,  {Li} S.,   {Gu} Q.,
  2020, \mn@doi [\apj] {10.3847/1538-4357/ab9b75}, \href
  {https://ui.adsabs.harvard.edu/abs/2020ApJ...897..162G} {897, 162}

\bibitem[\protect\citeauthoryear{{Hao}, {Shi}, {Chen}, {Xia}, {Gu}, {Guo}, {Yu}
   \& {Li}}{{Hao} et~al.}{2019}]{Hao_et_al.(2019)}
{Hao} C.-N.,  {Shi} Y.,  {Chen} Y.,  {Xia} X.,  {Gu} Q.,  {Guo} R.,  {Yu} X.,
  {Li} S.,  2019, \mn@doi [\apjl] {10.3847/2041-8213/ab42e5}, \href
  {https://ui.adsabs.harvard.edu/abs/2019ApJ...883L..36H} {883, L36}

\bibitem[\protect\citeauthoryear{{Hoffman}, {Cox}, {Dutta}  \&
  {Hernquist}}{{Hoffman} et~al.}{2010}]{Hoffman_et_al.(2010)}
{Hoffman} L.,  {Cox} T.~J.,  {Dutta} S.,   {Hernquist} L.,  2010, \mn@doi
  [\apj] {10.1088/0004-637X/723/1/818}, \href
  {https://ui.adsabs.harvard.edu/abs/2010ApJ...723..818H} {723, 818}

\bibitem[\protect\citeauthoryear{{Johansson}, {Naab}  \& {Burkert}}{{Johansson}
  et~al.}{2009}]{Johansson_et_al.(2009a)}
{Johansson} P.~H.,  {Naab} T.,   {Burkert} A.,  2009, \mn@doi [\apj]
  {10.1088/0004-637X/690/1/802}, \href
  {https://ui.adsabs.harvard.edu/abs/2009ApJ...690..802J} {690, 802}

\bibitem[\protect\citeauthoryear{{Krajnovi{\'c}} et~al.,}{{Krajnovi{\'c}}
  et~al.}{2011}]{Krajnovic_et_al.(2011)}
{Krajnovi{\'c}} D.,  et~al., 2011, \mn@doi [\mnras]
  {10.1111/j.1365-2966.2011.18560.x}, \href
  {https://ui.adsabs.harvard.edu/abs/2011MNRAS.414.2923K} {414, 2923}

\bibitem[\protect\citeauthoryear{{Li}, {Mao}, {Emsellem}, {Xu}, {Springel}  \&
  {Krajnovi{\'c}}}{{Li} et~al.}{2018}]{Li_et_al.(2018a)}
{Li} H.,  {Mao} S.,  {Emsellem} E.,  {Xu} D.,  {Springel} V.,   {Krajnovi{\'c}}
  D.,  2018, \mn@doi [\mnras] {10.1093/mnras/stx2374}, \href
  {https://ui.adsabs.harvard.edu/abs/2018MNRAS.473.1489L} {473, 1489}

\bibitem[\protect\citeauthoryear{{Lu} et~al.,}{{Lu}
  et~al.}{2021a}]{Lu_et_al.(2021b)}
{Lu} S.,  et~al., 2021a, arXiv e-prints, \href
  {https://ui.adsabs.harvard.edu/abs/2021arXiv210906197L} {p. arXiv:2109.06197}

\bibitem[\protect\citeauthoryear{{Lu} et~al.,}{{Lu}
  et~al.}{2021b}]{Lu_et_al.(2021a)}
{Lu} S.,  et~al., 2021b, \mn@doi [\mnras] {10.1093/mnras/stab497}, \href
  {https://ui.adsabs.harvard.edu/abs/2021MNRAS.503..726L} {503, 726}

\bibitem[\protect\citeauthoryear{{Marinacci} et~al.,}{{Marinacci}
  et~al.}{2018}]{Marinacci_et_al.(2018)}
{Marinacci} F.,  et~al., 2018, \mn@doi [\mnras] {10.1093/mnras/sty2206}, \href
  {http://adsabs.harvard.edu/abs/2018MNRAS.480.5113M} {480, 5113}

\bibitem[\protect\citeauthoryear{{Mihos} \& {Hernquist}}{{Mihos} \&
  {Hernquist}}{1996}]{Mihos_et_al.(1996)}
{Mihos} J.~C.,  {Hernquist} L.,  1996, \mn@doi [\apj] {10.1086/177353}, \href
  {https://ui.adsabs.harvard.edu/abs/1996ApJ...464..641M} {464, 641}

\bibitem[\protect\citeauthoryear{{Naab} \& {Burkert}}{{Naab} \&
  {Burkert}}{2003}]{Naab_et_al.(2003)}
{Naab} T.,  {Burkert} A.,  2003, \mn@doi [\apj] {10.1086/378581}, \href
  {https://ui.adsabs.harvard.edu/abs/2003ApJ...597..893N} {597, 893}

\bibitem[\protect\citeauthoryear{{Naiman} et~al.,}{{Naiman}
  et~al.}{2018}]{Naiman_et_al.(2018)}
{Naiman} J.~P.,  et~al., 2018, \mn@doi [\mnras] {10.1093/mnras/sty618}, \href
  {http://adsabs.harvard.edu/abs/2018MNRAS.477.1206N} {477, 1206}

\bibitem[\protect\citeauthoryear{{Nelson} et~al.,}{{Nelson}
  et~al.}{2015}]{Nelson_et_al.(2015)}
{Nelson} D.,  et~al., 2015, \mn@doi [Astronomy and Computing]
  {10.1016/j.ascom.2015.09.003}, \href
  {http://adsabs.harvard.edu/abs/2015A%26C....13...12N} {13, 12}

\bibitem[\protect\citeauthoryear{{Nelson} et~al.,}{{Nelson}
  et~al.}{2018}]{Nelson_et_al.(2018)}
{Nelson} D.,  et~al., 2018, \mn@doi [\mnras] {10.1093/mnras/stx3040}, \href
  {http://adsabs.harvard.edu/abs/2018MNRAS.475..624N} {475, 624}

\bibitem[\protect\citeauthoryear{{Nelson} et~al.,}{{Nelson}
  et~al.}{2019a}]{Nelson_et_al.(2019a)}
{Nelson} D.,  et~al., 2019a, \mn@doi [Computational Astrophysics and Cosmology]
  {10.1186/s40668-019-0028-x}, \href
  {https://ui.adsabs.harvard.edu/abs/2019ComAC...6....2N} {6, 2}

\bibitem[\protect\citeauthoryear{{Nelson} et~al.,}{{Nelson}
  et~al.}{2019b}]{Nelson_et_al.(2019b)}
{Nelson} D.,  et~al., 2019b, \mn@doi [\mnras] {10.1093/mnras/stz2306}, \href
  {https://ui.adsabs.harvard.edu/abs/2019MNRAS.490.3234N} {490, 3234}

\bibitem[\protect\citeauthoryear{{Nelson} et~al.,}{{Nelson}
  et~al.}{2021}]{Nelson_et_al.(2021)}
{Nelson} E.~J.,  et~al., 2021, arXiv e-prints, \href
  {https://ui.adsabs.harvard.edu/abs/2021arXiv210112212N} {p. arXiv:2101.12212}

\bibitem[\protect\citeauthoryear{{Pillepich} et~al.,}{{Pillepich}
  et~al.}{2018a}]{Pillepich_et_al.(2018a)}
{Pillepich} A.,  et~al., 2018a, \mn@doi [\mnras] {10.1093/mnras/stx2656}, \href
  {http://adsabs.harvard.edu/abs/2018MNRAS.473.4077P} {473, 4077}

\bibitem[\protect\citeauthoryear{{Pillepich} et~al.,}{{Pillepich}
  et~al.}{2018b}]{Pillepich_et_al.(2018b)}
{Pillepich} A.,  et~al., 2018b, \mn@doi [\mnras] {10.1093/mnras/stx3112}, \href
  {https://ui.adsabs.harvard.edu/abs/2018MNRAS.475..648P} {475, 648}

\bibitem[\protect\citeauthoryear{{Pillepich} et~al.,}{{Pillepich}
  et~al.}{2019}]{Pillepich_et_al.(2019)}
{Pillepich} A.,  et~al., 2019, \mn@doi [\mnras] {10.1093/mnras/stz2338}, \href
  {https://ui.adsabs.harvard.edu/abs/2019MNRAS.490.3196P} {490, 3196}

\bibitem[\protect\citeauthoryear{{Pop}, {Pillepich}, {Amorisco}  \&
  {Hernquist}}{{Pop} et~al.}{2018}]{Pop_et_al.(2018)}
{Pop} A.-R.,  {Pillepich} A.,  {Amorisco} N.~C.,   {Hernquist} L.,  2018,
  \mn@doi [\mnras] {10.1093/mnras/sty1932}, \href
  {https://ui.adsabs.harvard.edu/abs/2018MNRAS.480.1715P} {480, 1715}

\bibitem[\protect\citeauthoryear{{Pulsoni}, {Gerhard}, {Arnaboldi},
  {Pillepich}, {Nelson}, {Hernquist}  \& {Springel}}{{Pulsoni}
  et~al.}{2020}]{Pulsoni_et_al.(2020)}
{Pulsoni} C.,  {Gerhard} O.,  {Arnaboldi} M.,  {Pillepich} A.,  {Nelson} D.,
  {Hernquist} L.,   {Springel} V.,  2020, \mn@doi [\aap]
  {10.1051/0004-6361/202038253}, \href
  {https://ui.adsabs.harvard.edu/abs/2020A&A...641A..60P} {641, A60}

\bibitem[\protect\citeauthoryear{{Pulsoni}, {Gerhard}, {Arnaboldi},
  {Pillepich}, {Rodriguez-Gomez}, {Nelson}, {Hernquist}  \&
  {Springel}}{{Pulsoni} et~al.}{2021}]{Pulsoni_et_al.(2021)}
{Pulsoni} C.,  {Gerhard} O.,  {Arnaboldi} M.,  {Pillepich} A.,
  {Rodriguez-Gomez} V.,  {Nelson} D.,  {Hernquist} L.,   {Springel} V.,  2021,
  \mn@doi [\aap] {10.1051/0004-6361/202039166}, \href
  {https://ui.adsabs.harvard.edu/abs/2021A&A...647A..95P} {647, A95}

\bibitem[\protect\citeauthoryear{{Rodriguez-Gomez} et~al.,}{{Rodriguez-Gomez}
  et~al.}{2015}]{Rodriguez-Gomez_et_al.(2015)}
{Rodriguez-Gomez} V.,  et~al., 2015, \mn@doi [\mnras] {10.1093/mnras/stv264},
  \href {https://ui.adsabs.harvard.edu/abs/2015MNRAS.449...49R} {449, 49}

\bibitem[\protect\citeauthoryear{{Rodriguez-Gomez} et~al.,}{{Rodriguez-Gomez}
  et~al.}{2017}]{Rodriguez-Gomez_et_al.(2017)}
{Rodriguez-Gomez} V.,  et~al., 2017, \mn@doi [\mnras] {10.1093/mnras/stx305},
  \href {https://ui.adsabs.harvard.edu/abs/2017MNRAS.467.3083R} {467, 3083}

\bibitem[\protect\citeauthoryear{{Spilker} et~al.,}{{Spilker}
  et~al.}{2018}]{Spilker_et_al.(2018)}
{Spilker} J.,  et~al., 2018, \mn@doi [\apj] {10.3847/1538-4357/aac438}, \href
  {https://ui.adsabs.harvard.edu/abs/2018ApJ...860..103S} {860, 103}

\bibitem[\protect\citeauthoryear{{Springel}}{{Springel}}{2010}]{Springel(2010)}
{Springel} V.,  2010, \mn@doi [\mnras] {10.1111/j.1365-2966.2009.15715.x},
  \href {http://adsabs.harvard.edu/abs/2010MNRAS.401..791S} {401, 791}

\bibitem[\protect\citeauthoryear{{Springel}, {White}, {Tormen}  \&
  {Kauffmann}}{{Springel} et~al.}{2001}]{Springel_et_al.(2001)}
{Springel} V.,  {White} S.~D.~M.,  {Tormen} G.,   {Kauffmann} G.,  2001,
  \mn@doi [\mnras] {10.1046/j.1365-8711.2001.04912.x}, \href
  {http://adsabs.harvard.edu/abs/2001MNRAS.328..726S} {328, 726}

\bibitem[\protect\citeauthoryear{{Springel} et~al.,}{{Springel}
  et~al.}{2018}]{Springel_et_al.(2018)}
{Springel} V.,  et~al., 2018, \mn@doi [\mnras] {10.1093/mnras/stx3304}, \href
  {http://adsabs.harvard.edu/abs/2018MNRAS.475..676S} {475, 676}

\bibitem[\protect\citeauthoryear{{Stewart}}{{Stewart}}{2017}]{Stewart(2017)}
{Stewart} K.~R.,  2017, {Gas Accretion and Angular Momentum}.
p.~249, \mn@doi{10.1007/978-3-319-52512-9_11}

\bibitem[\protect\citeauthoryear{{Stoughton} et~al.,}{{Stoughton}
  et~al.}{2002}]{Stoughton_et_al.(2002)}
{Stoughton} C.,  et~al., 2002, \mn@doi [\aj] {10.1086/324741}, \href
  {http://adsabs.harvard.edu/abs/2002AJ....123..485S} {123, 485}

\bibitem[\protect\citeauthoryear{{Tacchella} et~al.,}{{Tacchella}
  et~al.}{2015}]{Tacchella_et_al.(2015)}
{Tacchella} S.,  et~al., 2015, \mn@doi [Science] {10.1126/science.1261094},
  \href {https://ui.adsabs.harvard.edu/abs/2015Sci...348..314T} {348, 314}

\bibitem[\protect\citeauthoryear{{Tacchella}, {Dekel}, {Carollo}, {Ceverino},
  {DeGraf}, {Lapiner}, {Mandelker}  \& {Primack}}{{Tacchella}
  et~al.}{2016}]{Tacchella_et_al.(2016)}
{Tacchella} S.,  {Dekel} A.,  {Carollo} C.~M.,  {Ceverino} D.,  {DeGraf} C.,
  {Lapiner} S.,  {Mandelker} N.,   {Primack} J.~R.,  2016, \mn@doi [\mnras]
  {10.1093/mnras/stw303}, \href
  {https://ui.adsabs.harvard.edu/abs/2016MNRAS.458..242T} {458, 242}

\bibitem[\protect\citeauthoryear{{Tumlinson}, {Peeples}  \& {Werk}}{{Tumlinson}
  et~al.}{2017}]{Tumlinson_et_al.(2017)}
{Tumlinson} J.,  {Peeples} M.~S.,   {Werk} J.~K.,  2017, \mn@doi [\araa]
  {10.1146/annurev-astro-091916-055240}, \href
  {https://ui.adsabs.harvard.edu/abs/2017ARA&A..55..389T} {55, 389}

\bibitem[\protect\citeauthoryear{{Vogelsberger}, {Genel}, {Sijacki}, {Torrey},
  {Springel}  \& {Hernquist}}{{Vogelsberger}
  et~al.}{2013}]{Vogelsberger_et_al.(2013)}
{Vogelsberger} M.,  {Genel} S.,  {Sijacki} D.,  {Torrey} P.,  {Springel} V.,
  {Hernquist} L.,  2013, \mn@doi [\mnras] {10.1093/mnras/stt1789}, \href
  {https://ui.adsabs.harvard.edu/abs/2013MNRAS.436.3031V} {436, 3031}

\bibitem[\protect\citeauthoryear{{Vogelsberger} et~al.,}{{Vogelsberger}
  et~al.}{2014a}]{Vogelsberger_et_al.(2014b)}
{Vogelsberger} M.,  et~al., 2014a, \mn@doi [\mnras] {10.1093/mnras/stu1536},
  \href {https://ui.adsabs.harvard.edu/abs/2014MNRAS.444.1518V} {444, 1518}

\bibitem[\protect\citeauthoryear{{Vogelsberger} et~al.,}{{Vogelsberger}
  et~al.}{2014b}]{Vogelsberger_et_al.(2014a)}
{Vogelsberger} M.,  et~al., 2014b, \mn@doi [\nat] {10.1038/nature13316}, \href
  {https://ui.adsabs.harvard.edu/abs/2014Natur.509..177V} {509, 177}

\bibitem[\protect\citeauthoryear{{Vogelsberger} et~al.,}{{Vogelsberger}
  et~al.}{2018}]{Vogelsberger_et_al.(2018)}
{Vogelsberger} M.,  et~al., 2018, \mn@doi [\mnras] {10.1093/mnras/stx2955},
  \href {https://ui.adsabs.harvard.edu/abs/2018MNRAS.474.2073V} {474, 2073}

\bibitem[\protect\citeauthoryear{{Vogelsberger}, {Marinacci}, {Torrey}  \&
  {Puchwein}}{{Vogelsberger} et~al.}{2020a}]{Vogelsberger_et_al.(2020a)}
{Vogelsberger} M.,  {Marinacci} F.,  {Torrey} P.,   {Puchwein} E.,  2020a,
  \mn@doi [Nature Reviews Physics] {10.1038/s42254-019-0127-2}, \href
  {https://ui.adsabs.harvard.edu/abs/2020NatRP...2...42V} {2, 42}

\bibitem[\protect\citeauthoryear{{Vogelsberger} et~al.,}{{Vogelsberger}
  et~al.}{2020b}]{Vogelsberger_et_al.(2020b)}
{Vogelsberger} M.,  et~al., 2020b, \mn@doi [\mnras] {10.1093/mnras/staa137},
  \href {https://ui.adsabs.harvard.edu/abs/2020MNRAS.492.5167V} {492, 5167}

\bibitem[\protect\citeauthoryear{{Wang}, {Xu}, {Lu}, {Cai}, {Xiang}, {Mao},
  {Springel}  \& {Hernquist}}{{Wang} et~al.}{2021}]{Wang_et_al.(2021)}
{Wang} S.,  {Xu} D.,  {Lu} S.,  {Cai} Z.,  {Xiang} M.,  {Mao} S.,  {Springel}
  V.,   {Hernquist} L.,  2021, arXiv e-prints, \href
  {https://ui.adsabs.harvard.edu/abs/2021arXiv210906200W} {p. arXiv:2109.06200}

\bibitem[\protect\citeauthoryear{{Weinberger} et~al.,}{{Weinberger}
  et~al.}{2017}]{Weinberger_et_al.(2017)}
{Weinberger} R.,  et~al., 2017, \mn@doi [\mnras] {10.1093/mnras/stw2944}, \href
  {https://ui.adsabs.harvard.edu/abs/2017MNRAS.465.3291W} {465, 3291}

\bibitem[\protect\citeauthoryear{{Weinberger} et~al.,}{{Weinberger}
  et~al.}{2018}]{Weinberger_et_al.(2018)}
{Weinberger} R.,  et~al., 2018, \mn@doi [\mnras] {10.1093/mnras/sty1733}, \href
  {https://ui.adsabs.harvard.edu/abs/2018MNRAS.479.4056W} {479, 4056}

\bibitem[\protect\citeauthoryear{{Xu}, {Springel}, {Sluse}, {Schneider},
  {Sonnenfeld}, {Nelson}, {Vogelsberger}  \& {Hernquist}}{{Xu}
  et~al.}{2017}]{Xu_et_al.(2017)}
{Xu} D.,  {Springel} V.,  {Sluse} D.,  {Schneider} P.,  {Sonnenfeld} A.,
  {Nelson} D.,  {Vogelsberger} M.,   {Hernquist} L.,  2017, \mn@doi [\mnras]
  {10.1093/mnras/stx899}, \href
  {http://adsabs.harvard.edu/abs/2017MNRAS.469.1824X} {469, 1824}

\bibitem[\protect\citeauthoryear{{Xu} et~al.,}{{Xu}
  et~al.}{2019}]{Xu_et_al.(2019)}
{Xu} D.,  et~al., 2019, \mn@doi [\mnras] {10.1093/mnras/stz2164}, \href
  {https://ui.adsabs.harvard.edu/abs/2019MNRAS.489..842X} {489, 842}

\bibitem[\protect\citeauthoryear{{Zeng}, {Wang}  \& {Gao}}{{Zeng}
  et~al.}{2021}]{Zeng_et_al.(2021)}
{Zeng} G.,  {Wang} L.,   {Gao} L.,  2021, \mn@doi [\mnras]
  {10.1093/mnras/stab2294}, \href
  {https://ui.adsabs.harvard.edu/abs/2021MNRAS.507.3301Z} {507, 3301}

\bibitem[\protect\citeauthoryear{{Zhang} et~al.,}{{Zhang}
  et~al.}{2019}]{Zhang_et_al.(2019)}
{Zhang} C.,  et~al., 2019, \mn@doi [\apjl] {10.3847/2041-8213/ab4ae4}, \href
  {https://ui.adsabs.harvard.edu/abs/2019ApJ...884L..52Z} {884, L52}

\bibitem[\protect\citeauthoryear{{Zhou}, {Li}, {Hao}, {Guo}, {Mo}  \&
  {Xia}}{{Zhou} et~al.}{2020}]{Zhou_et_al.(2020)}
{Zhou} S.,  {Li} C.,  {Hao} C.-N.,  {Guo} R.,  {Mo} H.,   {Xia} X.,  2020,
  arXiv e-prints, \href {https://ui.adsabs.harvard.edu/abs/2020arXiv201113749Z}
  {p. arXiv:2011.13749}

\bibitem[\protect\citeauthoryear{{de Vaucouleurs}}{{de
  Vaucouleurs}}{1948}]{de_Vaucouleurs(1948)}
{de Vaucouleurs} G.,  1948, Annales d'Astrophysique, \href
  {http://adsabs.harvard.edu/abs/1948AnAp...11..247D} {11, 247}

\makeatother
\end{thebibliography}



\label{lastpage}
\end{document}